\documentclass[12pt,PhD,singlespace]{mythesis}

\usepackage{amsmath,amssymb,amsfonts,mathrsfs,bm,cite}
\usepackage{graphicx,axodraw,epic,eepic,epsf}
\usepackage{FeynMan,myDef}

\begin{document}
\figurespagefalse
\tablespagefalse

\title{Quantization and High Energy Unitarity in Orbifold Theories}
\author{Lars Nilse}
\school{Physics and Astronomy}
\principaladviser{Dr. Apostolos Pilaftsis (Manchester University)}
\firstreader{Prof. Jeff Forshaw (Manchester University)}
\secondreader{Dr. Steve Abel (Durham University)}
\submitdate{May 2005}

\beforeabstract
\prefacesection{Abstract}
We study five-dimensional Yang-Mills theories compactified on an $S^1/\Zb_2$ orbifold. The fundamental Lagrangian naturally includes brane kinetic terms at the orbifold fixed points which are induced by quantum corrections of the bulk fields. The theories are quantized in the higher-dimensional $R_\xi$ gauges before compactification. Using Ward and Slavnov-Taylor identities, an all-order proof of a generalized equivalence theorem is presented. The theorem relates scattering amplitudes of longitudinal Kaluza-Klein gauge bosons to amplitudes of the corresponding scalar modes. Non-trivial sum rules among the fundamental couplings of the 4D effective theory lie at the heart of high energy unitarity cancellations. Using a novel coupled channel analysis, we derive an upper bound on the number of Kaluza-Klein modes from perturbative unitarity. The bound shows a very weak dependence on the size of the brane kinetic terms.

\afterabstract
\prefacesection{Acknowledgements}
I would like to thank my supervisor Apostolos Pilaftsis, who provided me with the opportunity to work in Beyond Standard Model Physics. His many good ideas and his careful approach made it a great pleasure to work with him. I learned a lot under his supervision.

I would also like to thank my collaborators Alexander M\"uck (Edinburgh University) and Reinhold R\"uckl (W\"urzburg University). Their great experience in extra dimensional theories and their questioning minds contributed greatly to the project.

I would like to thank Tom Underwood with whom I have shared an office over the last three years. The many interesting discussions with him, among them about his fascinating work on Resonant Leptogenesis, helped me to understand the wider picture of Beyond Standard Model Physics.
\afterpreface

\chapter{Introduction}
\label{introduction}

The Higgs of the Standard Model (SM) does not only break gauge symmetry and give rise to particle masses, but restores unitarity in high energy scattering processes. Any theory with the ambition to replace the electroweak symmetry breaking mechanism of the SM has to address all three of the above points satisfactorily.
 
The concept of extra spatial dimensions proves to be popular in modern physics, since it allows for new perspectives on a wide range of problems. We will present a short overview at the end of this section. In field theories with extra spatial dimensions, \textit{gauge symmetry breaking} can be realized either explicitly by boundary conditions~\cite{Scherk:1978ta, Scherk:1979zr, Cremmer:1979uq, Hebecker:2001jb} or dynamically via the Hosotani mechanism~\cite{Hosotani:1983xw, Hosotani:1983vn, Hosotani:1988bm, vonGersdorff:2002as}. In the latter case, the extra component of the higher dimensional gauge field acquires a vacuum expectation value and can therefore play the role of the Higgs field in the SM. \textit{Particle masses} arise naturally in these theories as a consequence of the Kaluza-Klein (KK) compactification. In this thesis, we will focus on the last of the above points and study in detail how the \textit{unitarity of the scattering matrix} is ensured in quantum field theories with extra dimensions. We will put particular emphasis on a careful quantization procedure.

The particular model of our study is five-dimensional (5D) Yang-Mills theory compactified on an $S^1/\Zb_2$ orbifold. The fundamental Lagrangian contains additional kinetic terms for the gauge fields that are localized at the orbifold fixed points. These so-called brane kinetic terms (BKT) act as counter terms that renormalize UV-infinite operators arising from quantum corrections of the bulk fields. They are therefore strictly necessary for a consistent formulation of the theory. We quantize the theory before compactification and thereby extend the approach in~\cite{Muck:2001yv} to orbifold theories with BKT. The resulting effective theory is free of any mixing terms, either between gauge and scalar sector or between different KK modes, and could be interpreted as a theory in which each individual KK gauge mode is gauge-fixed in the conventional 4D $R_\xi$ gauges.

The advantages of this 5D quantization method become apparent when studying the symmetries of the theory. The classical and quantized action of the higher-dimensional theory are invariant under standard gauge and BRS transformations, which in turn give rise to Ward and Slavnov-Taylor (ST) identities~\cite{Pokorski}. After KK reduction, these symmetries and identities have their counterparts in the effective theory. It is the set of these 4D Ward and Slavnov-Taylor identities that provides the basis of our high energy unitarity discussion. In an earlier study, the authors of~\cite{Chivukula:2001hz, Chivukula:2003kq} checked the equivalence theorem (ET) for particular processes at tree-level in orbifold theories without BKT. Here we are able to prove the more comprehensive Generalized Equivalence Theorem (GET) for arbitrary processes at all orders in orbifold theories with BKT. As we will see, the high energy unitarity relies on subtle cancellations between the modes of the entire KK tower.

The 5D quantum field theories we are starting from are non-renormalizable. Hence we can only treat them as effective theories~\cite{Georgi:AnnRev}; intermediates between a fundamental (most likely stringy) UV complete theory and an effective 4D theory from which we calculate our predictions for scattering amplitudes and decay widths. Perturbative unitarity puts limits on the size of partial wave amplitudes of scattering processes. We employ these limits to derive the energy scale, up to which the predictions of our effective theory can be trusted. This upper energy bound corresponds to a maximum KK number available in the effective theory. We find it to show a strikingly weak dependence on the strength of the localized terms. Even huge BKT are not able to screen the dynamics in the bulk. This is indeed plausible, since high energy unitarity probes length scales much smaller than the compactification radius and is therefore ignorant of any localized operators. 
\vspace{4mm}

\noindent
In the following, we give a very short overview of physics in higher-dimensional spacetimes in order to describe the context of our study. More complete introductions and reviews of extra dimensions can be found in~\cite{Hewett:AnnRev, Csaki:2004ay, Quiros:2003gg, Abel:PhysicsWorld, Randall:2002ie, Landsberg:2004mj}.

After Riemann~\cite{Riemann} had laid the mathematical foundations in the late nineteenth century, Nordstr\"om, Kaluza and Klein~\cite{Nordstroem, Kaluza:1921tu, Klein:1926fj} were the first to apply the concept of extra space dimensions to physics. Their early attempt to develop a unified theory of electromagnetism and gravity failed, and the idea of extra dimensions played no further role in the development of early twentieth century physics. It reemerged more than fifty years later within superstring theory, when Green and Schwarz~\cite{Green:1984sg, Green:1984ed} realized that 10 spacetime dimensions were needed for an anomalous-free, i.e. consistent, formulation of the theory. String theory was, and is, the only promising candidate for a quantum theory of gravity, and the concept of extra dimensions became widely accepted among theoretical physicists. With a 16-orders-of-magnitude gap between direct string theoretical predictions and the scale of their current experiments, experimental physicists remained less enthusiastic. That changed, when Antoniadis and independently Lykken~\cite{Antoniadis:1990ew, Lykken:1996fj, Antoniadis:1996hk} realized that compactification and string scale are not necessarily tied to the Planck scale. Earlier naive dimensional analyses had incorrectly assumed that all dimensionless parameters of the theory, such as the unified gauge coupling, had to be of order one. Suddenly, even $TeV^{-1}$ sized extra dimensions became imaginable. In our present discussion, we will distinguish between four different scenarios with large extra dimensions.

The first of them was proposed by \textit{Arkani-Hamed, Dimopoulos and Dvali (ADD)}~\cite{Arkani-Hamed:1998rs, Antoniadis:1998ig, Arkani-Hamed:1998nn} in 1998. The $d \geq 2$ extra dimensions in their framework are flat and compactified, i.e. finite. Gravity propagates in the entire bulk, whereas SM fields are restricted to a 3-brane. The large size $R$ of the extra dimensions is the key which enables the authors to solve, or rather evade, the gauge hierarchy problem. ADD argue that the fundamental higher-dimensional Planck scale $M^*_{\operatorname{Pl}}$ and our electroweak scale $M_{\operatorname{EW}}$ are identical. Four-dimensional gravity appears weak to us, since it is diluted by the volume of the compactified dimensions. A simple argument based on Gau\ss's law relates our Planck scale $M_{\operatorname{Pl}}$ in four dimension to the fundamental scale, $M^2_{\operatorname{Pl}} \sim R^d \, M^{* \; 2+d}_{\operatorname{Pl}}$.

A year later, \textit{Randall and Sundrum (RS)}~\cite{Randall:1999ee, Randall:1999vf} presented a radically different explanation of the hierarchy. Their model assumes a single warped extra dimension of finite or even infinite size. Gravity travels in the extra dimension and is localized on a 3-brane. The SM is stuck on a second brane some distance apart. The hierarchy arises this time from an exponential factor in the AdS background metric of the extra dimension.

In the scenario by \textit{Dvali, Gabadadze and Porrati (DGP)}~\cite{Dvali:2000hr, Deffayet:2001pu, Dvali:2002vf} the extra dimension is neither compact nor warped. A single 3-brane is located at $y=0$ in the flat, infinite extra dimension. The SM is again confined to the brane. The fundamental action $\int d^5 x \, ( \, M^{* \; 3}_{\operatorname{Pl}} \sqrt{| \mathcal{G} |} \mathcal{R} +  \delta(y) M^2_{\operatorname{Pl}} \sqrt{| g |} R \, )$ includes a BKT which is induced by loop corrections. $\mathcal{R}$ and $R$ are the Ricci scalars of the 5D metric $\mathcal{G}_{\mu \nu}$ and the induced 4D metric $g_{\mu \nu}(x)=\mathcal{G}_{\mu \nu}(x,y=0)$ respectively. The 5D Planck mass $M^*_{\operatorname{Pl}}$ and its 4D counterpart $M_{\operatorname{Pl}}$ are understood to be independent parameters of the theory. At short distances, the 4D scalar curvature term dominates and gravity is four-dimensional as observed. At very large distances the 5D term takes over, and contrary to the ADD scenario gravity appears to be weaker at cosmic distances. The crossover scale $M^2_{\operatorname{Pl}}/(2 M^{* \; 3}_{\operatorname{Pl}})$ is of the order of the present Hubble length. The leakage of gravity into the extra dimension at large distances can explain the observed acceleration of our Universe~\cite{Riess:1998cb, Perlmutter:1998np}. The introduction of a small cosmological constant, and therefore another hierarchy, is consequently not necessary.

The fourth scenario, known as \textit{Universal Extra Dimensions (UED)}~\cite{Appelquist:2000nn}, was put forward by Appelquist, Cheng and Dobrescu in 2000. It is the one we are going to work in. One or more flat extra dimensions are compactified on a mani- or orbifold. Unlike in ADD or RS, gravity is absent and the model has consequently nothing to say about the hierarchy between the electroweak and the Planck scale. In the most simple version of the scenario, all of the SM fields are allowed to propagate in the bulk; in this sense the extra dimensions are 'universal'. The SM in ADD, RS and DGP scenarios is fixed to the 3-brane and remains four-dimensional. In UED on the other hand, every single field is aware of the extra dimensions. It is this fact that allows UED to address a very wide spectrum of problems: Totally new gauge and SUSY breaking mechanisms become possible~\cite{Quiros:2003gg, Hebecker:2001jb}. The fermion mass hierarchy, proton life time~\cite{Arkani-Hamed:1999dc, Kawamura:2000ev, Buchmuller:2004eg, Alciati:2005ur} or the number of fermion generations~\cite{Dobrescu:2001ae} can be understood. Electroweak physics~\cite{Csaki:2003dt, Csaki:2003zu, Csaki:2003sh, Barbieri:2003pr, Gabriel:2004ua} has no longer to rely on the Higgs mechanism. In the 4D SM, gauge couplings run logarithmically in the energy scale. Dienes, Dudas and Gherghetta (DDG)~\cite{Dienes:1998vh, Dienes:1998vg, Dienes:1999sz} pointed out that couplings in higher-dimensional theories follow a power law. The couplings unify earlier and the hierarchy between the electroweak and the GUT scale is consequently reduced. Even dark matter~\cite{Munoz:2003gx, Gaitskell:AnnRev, Cheng:2002ej, Servant:2002aq, Servant:2002hb, Hooper:2002gs, Bertone:2002ms}, dark energy and quintessence~\cite{Peloso:2003nv, Byrne:2004pc, Matsuda:2004ci} can be viewed from a new perspective.
\vspace{4mm}

\noindent
If large extra dimensions are not merely a possibility but reality, evidence of them is most likely to show up in one of the following three types of experiments: First there are classical \textit{collider searches}~\cite{Landsberg:2004mj}. Future machines such as the Large Hadron Collider (LHC) or the International Linear Collider (ILC)~\cite{Weiglein:2004hn, Dawson:AnnRev} might be able to probe energy scales characteristic for extra dimensions. References~\cite{Antoniadis:2000vd, Davoudiasl:1999ni, Ghosh:1999ex} study for example signatures for ADD scenarios. Signatures for UED might be very similar to the ones for SUSY, and linear collider studies are likely to be necessary to distinguish between the two~\cite{Battaglia:2005zf}. On the other hand, UED evidence might be hidden in old-fashioned  precision measurements~\cite{Buras:2002ej, Oliver:2002wn, Muck:2003kx}.

A further promising approach are sub-millimeter \textit{Inverse Square Law (ISL)} tests of gravity~\cite{Adelberger:AnnRev}. The most sensitive of these tests are torsion pendulum experiments, for example the one currently conducted by the E\"ot-Wash group~\cite{Hoyle:2004cw} at Washington University. New planar geometries~\cite{Long:Nature} or novel concepts using neutrons in the gravitational field of the earth~\cite{Nesvizhevsky:2004qb} might push the limits further in the years to come.

Other severe constraints on the size of extra dimensions arise from \textit{astrophysics and cosmology}~\cite{Hewett:AnnRev}. Possible implications include core cooling of supernovae, see study of SN1987A~\cite{Hanhart:2000er, Hirata:1987hu, Bionta:1987qt}, a cosmic diffuse gamma-ray background~\cite{Hannestad:2001jv, Hannestad:2001nq}, neutron star heat excess~\cite{Hannestad:2001xi}, overclosure of the Universe~\cite{Hall:1999mk} or a very early matter-dominated phase in the evolution of the Universe~\cite{Fairbairn:2001ct}.
\vspace{4mm}

\noindent
Finally, we would like to comment on the origin of the brane localized terms that are to appear in our fundamental Lagrangian (\ref{Lag5DYM}). BKTs are first discussed in the context of localized gravity in the DGP framework~\cite{Dvali:2001gm, Dvali:2000rx}. The concept is subsequently applied to SM fields in UED~\cite{Georgi:2000ks, Cheng:2002iz}. The authors calculate quantum corrections to self-energies in the higher-dimensional theory. Due to the orbifold symmetry, Lorentz invariance of the higher-dimensional theory is broken, which is reflected in the form of the propagators. Take for example the propagator of an even/odd scalar.

\begin{equation*}
D(p,p_5;q,q_5) \, = \, \frac{i}{p^2-p_5^2} \, \frac{\delta_{p_5, \, q_5} \pm \delta_{-p_5, \, q_5}}{2} \, \delta^{(4)}(p-q)
\end{equation*}

\noindent
It has got one contribution that no longer preserves the discrete component of the five-momentum, $p_5 \neq q_5$. Characteristic loop corrections involve two of these propagators. After integration and summation over the internal momenta, the self-energy correction splits into two parts. One contributes to the renormalization of the bulk fields. A second term is localized at the orbifold fixed points and was not present in the original Lagrangian. The detailed calculations are described in~\cite{Georgi:2000ks, Cheng:2002iz}.
\vspace{4mm}

\noindent
The thesis is structured as follows. Chapter~\ref{WardSlavnovTaylor} starts with a brief review of the derivation of Ward and ST identities in 4D Yang-Mills theories. It is followed by our study of 5D orbifold theories with BKT, starting with their quantization and compactification. The orthonormal basis used in this process is derived in Appendix~\ref{MassEigenmodeExpansion}. Appendix~\ref{Products} develops tools that prove to be helpful when deriving Ward and ST identities for the 4D effective theory. Based on these identities, Chapter~\ref{GET} presents a proof of the GET and studies particular examples. The calculations in this chapter make use of sum rules that are derived in Appendix~\ref{summation}. Chapter~\ref{HighEnergyUnitarityBounds} presents the derivation of upper bounds on the KK modes from perturbative unitarity, after which some final conclusions are drawn in Chapter~\ref{Conclusions}. The results of this thesis have been published in~\cite{Muck:2004br}.


\chapter{Ward and Slavnov-Taylor identities}
\label{WardSlavnovTaylor}

Let us start by reviewing the derivation of Ward and Slavnov-Taylor identities in four-dimensional Yang-Mills theories, such as standard QCD. We will fix our notation and discuss the basic principles, before applying them to compactified five-dimensional Yang-Mills theories in the following section.

\section{4D Yang-Mills theories}
\label{4DYM}

The Lagrangian of 4D Yang-Mills theories quantized in the $R_{\xi}$ gauges is given by

\begin{equation}
\label{Lag4DYM}
\Lag_{\YM} = - \frac{1}{4} F^a_{\mu \nu} F^{a \; \mu \nu} + \Lag_{\GF} + \Lag_{\FP} \: ,
\end{equation}

\noindent
where  $F^a_{\mu \nu}   =  \partial_{\mu} A^a_{\nu}  -  \partial_{\nu}
A^a_{\mu}+g f^{abc} A^b_{\mu} A^c_{\nu}$  denotes the field-strength tensor of the gluon field $A^a_{\mu}$. The gauge-fixing term $\Lag_{\GF}$ and the associated ghost term $\Lag_{\FP}$ are given by 

\begin{align}
\Lag_{\GF}  &=  -\, \frac{1}{2  \xi} \big(F[A^a_{\mu}]\big)^2 = - \frac{1}{2 \xi}\;  (\partial^{\mu}    A^a_{\mu})^2 \: ,\\
\Lag_{\FP} &= \overline{c}^a\, \frac{\delta  F[A^a_{\mu}]}{\delta \theta^b} \,  c^b = \overline{c}^a\, \big(\delta^{ab} \partial^2\:-\: g f^{abc}\, \partial^{\mu} A^c_{\mu}\,\big)\,c^b \: .
\end{align}

\noindent
The classical part of the Lagrangian $\Lag_{\YM}$, i.e. the first term in (\ref{Lag4DYM}), is invariant under the usual gauge transformations

\begin{equation}
\label{gauge4D}
\delta A^a_{\mu} = D^{ab}_{\mu}\, \theta^b = \big(\, \delta^{ab} \partial_{\mu}\: -\: g f^{abc}\, A^c_{\mu}\,\big)\, \theta^b \: .
\end{equation}

\noindent
The tree-level effective action is identical to the classical action itself, $\Gamma[A^a_{\mu}] = -\frac{1}{4}  \int d^4x \;  F^a_{\mu \nu} F^{a \, \mu \nu}$, and hence invariant under the above transformations too. From this invariance we can immediately derive the master Ward identity for 4D Yang-Mills theories.

\begin{gather}
\Gamma [A^a_{\mu}] =  \Gamma [A^a_\mu  +  \delta A^a_\mu] \\
\label{masterWard4DYM}
\partial_{\mu} \frac{\delta \Gamma}{\delta A^a_{\mu}} - g f^{abc}\; \frac{\delta \Gamma}{\delta A^b_{\mu}} A^c_{\mu} = 0
\end{gather}

\noindent
Functionally differentiating the identity with respect to the gluon fields $A^a_{\mu}$ and finally setting all fields to zero, we arrive at particular Ward identities that relate the one-particle irreducible (1PI) n-point functions of the theory. Since we started off from the \textit{tree-level} effective action $\Gamma$, we derive relations between the fundamental interactions of the theory. Let us follow a particular example. Functionally differentiating (\ref{masterWard4DYM}) with respect to $A^d_\nu (y)$, we find

\begin{equation*}
\begin{split}
\partial_\mu \frac{\delta \Gamma}{\delta A^a_\mu (x) \, \delta A^d_\nu (y)} &- g f^{abc} \frac{\delta \Gamma}{\delta A^b_\mu (x) \, \delta A^d_\nu (y)} \; A^c_\mu (x)\\
&- g f^{abc} \frac{\delta \Gamma}{\delta A^b_\mu (x)} \; \delta^{cd} \, g^\nu_\mu \, \delta^{(4)}(x-y) =0 \; .
 \end{split}
\end{equation*}

\noindent
A further differentiation with respect to $A^e_\rho (z)$ results in an equation involving four terms. Setting all of the fields equal to zero, the term proportional to $A^c_\mu (x)$ drops out. Introducing the notation $G^{ab \cdots}_{\mu \nu \cdots} (x,y, \cdots) = \delta \Gamma / [\delta A^a_\mu (x) \, \delta A^b_\nu (y) \cdots]$ for the 1PI Greens functions, we derive a Ward identity in position space.

\begin{equation*}
\begin{split}
\partial^\mu G^{ade}_{\mu \nu \rho} (x,y,z) &- g f^{abe} G^{bd}_{\rho \nu} (x,y) \; \delta^{(4)}(x-z)\\
&- g f^{abd} G^{be}_{\nu \rho}(x,z) \; \delta^{(4)} (x-y) =0
\end{split}
\end{equation*}

\noindent
Fourier transforming the result, we arrive at a final Ward identity that can graphically represented\footnote{We follow the convention, that all momenta flow into the vertices. Note that the diagrams on the right hand side (RHS) of (\ref{Ward4DYM1}) stand for the 2-point functions and not the propagators.} as follows.
\begin{equation}
\label{Ward4DYM1}
\begin{split}
-i k_{\mu} \BBB{a \; \mu}{b \; \nu}{c \; \rho}{\shift k}{p}{\shift q} \hspace{-3mm} = \ g f^{abd} \BB{d \; \nu}{c \; \rho}{\shift q}\: +\: g f^{acd} \BB{d \; \rho}{b \; \nu}{\shift p}\\ &\phantom{}\\ &\phantom{}
\end{split}
\end{equation}

\noindent
Functionally differentiating (\ref{masterWard4DYM}) with respect to a third gluon field $A^f_\sigma (u)$ before setting all fields to zero, we can derive a Ward identity relating the fundamental cubic and quartic couplings of the theory.

\begin{equation}
\begin{split}
-i k_{\mu} \BBBB{a \; \mu}{b \; \nu}{c \; \rho}{d \; \sigma}{k}{\shift p}{q}{\shift r} &=\  g f^{abe} \BBB{e \; \nu}{c \; \rho}{d \; \sigma}{k+p}{q}{\shift r}\\ &\phantom{}\\
+\ g f^{ace} &\BBB{e \; \rho}{b \; \nu}{d \; \sigma}{k+q}{p}{\shift r} + g f^{ade} \BBB{e \; \sigma}{b \; \nu}{c \; \rho}{k+r}{p}{\shift q}\\ 
&\phantom{}\\ &\phantom{}
\end{split}
\end{equation}

\noindent
Let us now turn to the quantized theory. After quantization the full Lagrangian $\Lag_{\YM}$ is no longer gauge-invariant, but it does retain an invariance under BRS transformations,

\begin{alignat}{2}
\delta A^a_{\mu} &= \omega \: s A^a_{\mu} &&= \omega \: D^{ab}_{\mu} c^b \nonumber \\
\label{BRS4DYM}
\delta c^a &= \omega \: s\, c^a  &&= -\; \omega \: \frac{g f^{abc}}{2} c^b c^c\\
\delta \overline{c}^a &= \omega \: s\, \overline{c}^a &&= \omega \: \frac{F[A^a_{\mu}]}{\xi} = \omega \: \frac{\partial^{\mu} A^a_{\mu}}{\xi} \: , \nonumber
\end{alignat}

\noindent
where $\omega$ is a small Grassmann parameter, i.e. $\omega^2=0$. The nilpotency of the BRS operator, $s^2 A^a_{\mu} = s^2 c^a = s^2 \bar{c}^a = 0$, ensures the unitarity of the physical S-matrix.

Following the standard path-integral quantization formalism~\cite{Pokorski}, we introduce the functionals $W$ and $Z$, which generate the full and connected Green's functions respectively.  In addition to the sources of gluons, ghosts and antighosts, $J^{a \mu}$, $\overline{D}^a$ and $D^a$, we allow for the terms $K^{a \mu}$ and $M^a$. These are the sources of operators $s A^a_{\mu}$  and $s c^a$, and will ensure that our final expression (\ref{STmaster}) will be a simple first order functional differential equation.

\begin{equation}
\label{WZ}
\begin{split}
W = e^{iZ}\, = \, &\int\! DA \, Dc \, D\bar{c} \, \exp \Big[\, 
i\int\! d^4x\, \big(\, \Lag_{\YM}\\
 &+ J^{a \mu} A^a_{\mu} +
\overline{D}^a c^a + \overline{c}^a D^a + K^{a \mu} s A^a_{\mu} + M^a
sc^a\, \big)\,\Big]
\end{split}
\end{equation}

\noindent
The BRS invariance of the action and path-integral measure implies again the invariance of the generating functionals.

\begin{equation}
Z[A^a_{\mu},c^a,\bar{c}^a] = Z[A^a_{\mu} + \delta A^a_{\mu}, c^a + \delta c^a, \bar{c}^a + \delta \bar{c}^a]
\end{equation}

\noindent
Substituting the explicit form (\ref{BRS4DYM}) of the BRS transformations into (\ref{WZ}), we can rewrite the transformation of the fields as a transformation of the sources instead\footnote{It now becomes clear, why there was no need to introduce a source for $s \bar{c}^a$ in the definition of the generating functionals. Our gauge fixing functional $F[A^a_{\mu}]=\partial^{\mu} A^a_{\mu}$ is linear in the gluon field, and after integration by parts the existing current $J^{a \, \mu}$ can be used. On the other hand, the first two transformations in (\ref{BRS4DYM}) are clearly quadratic in fields.}. The master ST identity for 4D Yang-Mills theories follows directly.

\begin{equation}
\begin{split}
Z[J^{a \mu}&, \overline{D}^a, D^a, K^{a \mu}, M^a] =\\
&Z \big[J^{a \mu} - \frac{\omega}{\xi} \, \partial^\mu D^a, \overline{D}^a, D^a, K^{a \mu} + \omega J^{a \mu}, M^a - \omega \overline{D}^a \big]
\end{split}
\end{equation}
 
\begin{equation}
\label{STmaster}
J^{a \mu} \frac{\delta Z}{\delta K^{a \mu}} - \overline{D}^a \frac{\delta Z}{\delta M^a}\ +\ \frac{1}{\xi}\, D^a\, \partial^{\mu} \frac{\delta Z}{\delta J^{a \mu}}\ =\ 0
\end{equation}

\noindent
Earlier in this section we encountered already the effective action $\Gamma$, but worked with it only at tree-level. Defining the full effective action as usual, we can derive a master ST identity for 1PI Green's functions\footnote{Note that a full renormalizability proof for 4D Yang-Mills theories~\cite{ZinnJustin, Bardeen:1978cz} is based on (\ref{STmaster1PI}) and the equation of motion for the antighosts,
\begin{equation*}
\partial^{\mu} \frac{\delta \Gamma}{\delta K^{a \mu}} + \frac{\delta \Gamma}{\delta \overline{c}^a} = 0 \; .
\end{equation*}}.

\begin{align}
\begin{split}
\Gamma[A^a_{\mu}, &c^a, \bar{c}^a, K^{a \mu}, M^a] =\\
&Z[J^{a \mu}, \overline{D}^a, D^a, K^{a \mu}, M^a] - \int d^4 x \big( J^{a \mu} A^a_{\mu} + \overline{D}^a c^a + \overline{c}^a D^a \big)
\end{split}\\
&\text{with} \quad A^a_{\mu} = \frac{\delta Z}{\delta J^{a \mu}}, \quad c^a = \frac{\delta Z}{\delta \overline{D}^a}, \quad \overline{c}^a = - \frac{\delta Z}{\delta D^a} \; , \nonumber
\end{align}

\begin{equation}
\label{STmaster1PI}
\frac{\delta \Gamma}{\delta A^a_{\mu}} \frac{\delta \Gamma}{\delta K^{a \mu}} + \frac{\delta \Gamma}{\delta c^a} \frac{\delta \Gamma}{\delta M^a} - \frac{1}{\xi} \frac{\delta \Gamma}{\delta \overline{c}^a} \partial^{\mu} A^a_{\mu} = 0
\end{equation}

\noindent
As before, functional differentiation of the master equations will lead to particular ST identities for connected/1PI Green's functions. But there is a more direct way of deriving specific ST identities. It is based on the observation that the BRS invariance of the generating functionals implies the invariance of the Green's functions themselves. Let us consider the following example.

\begin{equation}
s\, \langle 0 | T \overline{c}^a(x) A^b_{\nu}(y) A^c_{\rho}(z) | 0
\rangle = 0
\end{equation}

\noindent
The Green's functions are given as vacuum expectation values of time-ordered operator products. Applying the transformations (\ref{BRS4DYM}) to each of the fields, we arrive at the following expression.

\begin{equation}
\label{STexample}
\begin{split}
\frac{1}{\xi}\, \partial^{\mu}_{x}\, \langle 0 | T A^a_{\mu} (x) A^b_{\nu} (y) A^c_{\rho} (z) | 0 \rangle\:
&-\: \partial^{y}_{\nu} \langle 0 | T \overline{c}^a(x) c^b(y) A^c_{\rho}(z)| 0 \rangle\\
-\: \partial^{z}_{\rho} \langle 0 | T \overline{c}^a(x) A^b_{\nu} (y) c^c(z) | 0 \rangle &+\:  g f^{bde} \langle 0 | T \overline{c}^a(x) c^d(y) A^e_{\nu} (y) A^c_{\rho}(z) | 0 \rangle\\ 
&+\: g f^{cde} \langle 0 | T \overline{c}^a(x) A^b_{\nu}(y) c^d(z) A^e_{\rho}(z) | 0 \rangle\ =\ 0
\end{split}
\end{equation}

\noindent
Note that the last two terms are bilinear in fields at the same space-time point, $y$ and $z$ respectively. These terms do not have one-particle poles for external legs attached to these points, and hence do not contribute on-shell. Fourier transforming (\ref{STexample}) and fixing the gauge, $\xi = 1$, we derive a particular on-shell ST identity in momentum space. In graphical notation the identity has got the form below, where the shaded circles stand for all-order 1PI Green's functions.
\vspace{-2mm}
\begin{equation}
p^{\mu} \; \BBBblob{a \; \mu}{b \; \nu}{c \; \rho}{p}{q}{\shift r} \nshift \nshift - q_{\nu} \GGBblob{a}{b}{c \; \rho}{p}{q}{\shift r} \nshift \nshift - r_{\rho} \GBGblob{a}{b \; \nu}{c}{p}{q}{\shift r} \nshift \nshift = 0
\end{equation}
\vspace{1cm}

\noindent
Let us state two further examples in order to illustrate the technique~\cite{Pascual}. The BRS invariance of the following two Green's functions gives rise to the (not necessarily on-shell) ST identities below\footnote{Particular care has to be taken, when space-time derivatives and time-ordering (i.e. space-time step-functions) are interchanged. In some cases, this interchange results in important Schwinger terms, as for example on the RHS of (\ref{SchwingerExample}).}.

\begin{equation}
\label{SchwingerExample}
\begin{split}
s \, \langle 0 | T \, \bar{c}^a(x) \, &\partial^{\nu}_y A^b_{\nu} (y) | 0 \rangle = 0\\
p^{\mu} q^{\nu} \BBblob{a \; \mu}{b \; \nu} &= p^{\mu} q^{\nu} \BB{a \, \mu}{b \, \nu}{}
\end{split}
\end{equation}
\vspace{3mm}

\begin{equation}
\begin{split}
s \, \langle 0 | T \, &\bar{c}^a(x) \, \partial^{\nu}_y A^b_{\nu} (y) \, \partial^{\rho}_z A^b_{\rho} (z) \, \partial^{\sigma}_w A^b_{\sigma} (w) | 0 \rangle = 0\\
&k_{\mu} \, p_{\nu} \, q_{\rho} \, r_{\sigma} \nshift \BBBBblob{a \; \mu}{b \; \nu}{c \; \rho}{d \; \sigma}{k}{\shift p}{q}{\shift r} \nshift = 0\\ \phantom{}\\ \phantom{}\\
\end{split}
\end{equation}

\noindent
As we will see in the next sections, it is the 5D analog of these ST identities that lie at the heart of our proof of the GET for 5D orbifold theories.


\section{Compactified 5D Yang-Mills theories}
\label{5DYM}

We now move on to 5D Yang-Mills theories compactified on an $S^1/\Zb_2$ orbifold. Here and in the following, Lorentz indices in five dimensions will be denoted by capital Roman letters, while Greek letters are used for the four uncompactified dimensions. For space-time coordinates we use the notation $x^M=(x^{\mu},x^5 \equiv y)$, where $y \in (- \pi R, \pi R]$ stands for the extra dimension and $R$ for the compactification radius.

The orbifold $\Zb_2$ symmetry, $y \sim -y$, breaks Lorentz invariance of the compact dimension, i.e. momentum is not conserved at the fixed points of the orbifold symmetry, $y=0$ and $y=\pi R$. It is hence no surprise, that we find UV divergent operators at the orbifold fixed points when calculating quantum corrections~\cite{Georgi:2000ks}. For a consistent theory it is therefore \textit{necessary} to include localized counter-terms, the so-called brane kinetic terms (BKT), in the tree-level Lagrangian in order to absorb these infinities.

We will restrict ourselves to a single BKT at $y=0$. Our entire discussion carries over to the general case of two BKTs at $y=0$ and $y=\pi R$, but in this case we would have to deal with more complicated analytic expressions. The Lagrangian we are working with reads

\begin{equation}
\label{Lag5DYM}
\Lag_{\rm 5D\YM} (x,y) = - \frac{1}{4}\, \big[ 1\: +\: r_c\,\delta(y) \big] \, F^a_{MN} F^{a \; MN} + \Lag_{\rm 5D\GF} + \Lag_{\rm 5D\FP} \; ,
\end{equation}

\noindent
where $F^a_{MN}  = \partial_M A^a_N  - \partial_N A^a_M +  g_5 f^{abc} A^b_M A^c_N$ is the field-strength tensor of the 5D gluon $A^a_M$. The positive dimensionful coupling $r_c$ of the BKT is a free parameter of the theory, ultimately to be fixed by a UV completion of our field theory. The terms $\Lag_{\rm 5D\GF}$ and $\Lag_{\rm 5D\FP}$ are the 5D gauge-fixing and ghost term and will be discussed in detail later on.

We would like to include a massless gluon in the effective theory and choose the following parities for the components of the 5D gluon field $A^a_M$.

\begin{equation}
\label{parities}
\begin{split}
A^a_M(x,y) &= A^a_M(x, y+2 \pi R)\\ 
A^a_{\mu}(x,y) &= A^a_{\mu}(x,-y)\\ 
A^a_5(x,y) &= - A^a_5(x,-y)
\end{split}
\end{equation}

\noindent
We can now expand the components in terms of two complete sets of orthonormal functions, the even $f_n(y)$ and the odd $g_n(y)$.

\begin{equation}
\label{componentExpansion}
\begin{split}
A^a_\mu (x,y) &= \sum_{n=0}^{\infty} A^a_{(n) \mu}(x)\, f_n(y)\\
A^a_5 (x,y) &= \sum_{n=1}^{\infty} A^a_{(n) 5}(x)\, g_n(y)
\end{split}
\end{equation}

\noindent
Integrating the extra dimension $y$ out, we end up with an effective 4D theory. The coefficients $A^a_{(n) \mu}$ and $A^a_{(n) 5}$ are the so-called Kaluza-Klein (KK) modes. By demanding each of the KK modes to be a mass eigenstate of the effective theory, we can uniquely determine the analytic form of $f_n$ and $g_n$. In the case of vanishing BKT, $r_c \to 0$, we recover the standard Fourier expansion, with $f_n(y)$ and $g_n(y)$ equal to $\cos(n y / R)$ and $\sin(n y / R)$ up to a normalization constant. For $r_c \neq 0$ we describe the details of the derivation in Appendix~\ref{MassEigenmodeExpansion}. Here we only state the result.

\begin{align}
\label{fn}
f_n(y) =& \frac{N_n}{\sqrt{2^{\delta_{n,0}} \pi R} \; \cos m_n \pi R} \times \begin{cases}
\cos m_n ( y + \pi R) \; &\textrm{for} \quad -\pi R < y \leq 0\\
\cos m_n ( y - \pi R) \; &\textrm{for}\qquad 0 < y \leq \pi R
\end{cases}\\[3mm]
\label{gn}
g_n(y) =& \frac{N_n}{\sqrt{\pi R} \; \cos m_n \pi R} \times \begin{cases}
\sin m_n ( y + \pi R) \; &\textrm{for} \quad -\pi R < y < 0\\
\sin m_n ( y - \pi R) \; &\textrm{for}\qquad 0 < y \leq \pi R\\
0 &\textrm{for} \qquad y=0
\end{cases}\\[3mm]
&N_n^{-2} = 1 + \tilde{r}_c + \pi^2 R^2 \tilde{r}_c^2 m_n^2 \qquad \textrm{with} \quad \tilde{r}_c = \frac{r_c}{2 \pi R} \geq 0 
\end{align}

\noindent
The same relations that we are familiar with from sine and cosine still hold for the $f_n$ and $g_n$.

\begin{equation}
\begin{split}
\partial_5 f_n &= - m_n g_n\\
\partial_5 g_n &= m_n f_n
\end{split}
\end{equation}
 
\noindent
The mass of the KK gauge bosons $A^a_{(n) \mu}$ is given by the transcendental equation below. It can be solved numerically, as illustrated in Fig.~2.1.

\begin{equation}
\label{1BKTspectrum}
\frac{m_n r_c}{2} = - \tan m_n \pi R
\end{equation}

\noindent
The presence of the BKT decreases the masses slightly, and we find the spectrum to vary in the limits $(n-1/2)/R < m_n \leq n/R$ for $\infty > r_c \geq 0$ and $n \neq 0$. For vanishing BKT we recover the familiar equi-spaced KK tower $n/R$. Using the technique developed in Appendix~\ref{Products}, it is easy to derive the Feynman rules for the effective 4D theory. They are listed in Appendix~\ref{FeynmanRules}.

\begin{figure}[!ht]
\begin{center}
\parbox{100mm}{
\begin{center}
\includegraphics[width=0.7\textwidth]{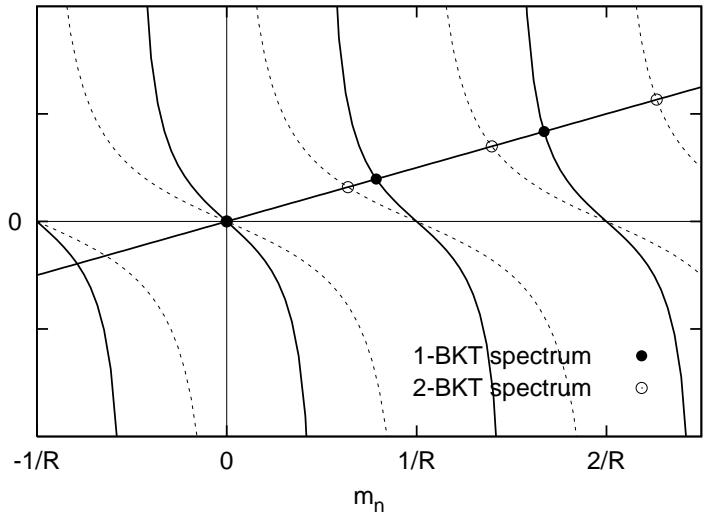}
\caption{Mass spectrum $m_n$ of the effective 4D theory for one (\ref{1BKTspectrum}) and two (\ref{2BKTspectrum}) BKT in the 5D Lagrangian.}
\end{center}}
\end{center}
\label{spectra}
\end{figure}

The orbifold symmetry acts only on space-time and leaves the invariance of the classical part of Lagrangian $\Lag_{\rm 5D\YM}$ under 5D gauge transformations totally unaffected.

\begin{equation}
\label{gauge5D}
\delta A^a_M = D^{ab}_M\, \theta^b = \big(\, \delta^{ab} \partial_M - g_5 f^{abc} A^c_M\,\big)\, \theta^b
\end{equation}

\noindent
As previously done for the components of the 5D gauge field, we can expand the gauge parameter $\theta^a(x,y)$ in terms of orthonormal functions\footnote{In order for (\ref{gauge5D}) to be consistent, an even $A^a_{\mu}$ implies $\theta^a$ to be even and hence $A^a_5$ to be odd. Demanding a massless gluon in the effective theory fixes therefore the parities in (\ref{parities}) uniquely.}.

\begin{equation}
\theta^a(x,y) = \sum_{n=0}^{\infty} \theta^a_{(n)}(x) \, f_n(y)
\end{equation}

\noindent
After integration over the compact dimension $y$, we find the effective gauge transformation for the KK modes,

\begin{equation}
\label{gaugeKK4D}
\begin{split}
\delta A^a_{(n) \mu} &= \partial_{\mu} \theta^a_{(n)} - g f^{abc} \sum_{m, l=0}^{\infty} \sqrt{2}^{\;
-1-\delta_{n,0}-\delta_{m,0}-\delta_{l,0}} \; \theta^b_{(m)}\, A^c_{(l) \mu}\, \Delta_{n,l,m}\\
\delta A^a_{(n) 5} &= - m_n \theta^a_{(n)} - g f^{abc} \sum_{m=0, \; l=1}^{\infty} \sqrt{2}^{\; -1-\delta_{m,0}} \; \theta^b_{(m)}\, A^c_{(l) 5} \, \tilde{\Delta}_{n,l,m} \; ,
\end{split}
\end{equation}

\noindent
where we introduced the dimensionless coupling constant $g = g_5 / \sqrt{2 \pi R}$. Expressions  for $\Delta_{k,l,n}$ and $\tilde{\Delta}_{k,l,n}$ can be found in Appendix~\ref{Products}. In the limit $r_c \to 0$, we find $\Delta_{k,l,n} \to \delta_{k,l,n}$, $\tilde{\Delta}_{k,l,n} \to \tilde{\delta}_{k,l,n}$, and the above transformations reduce to the ones familiar from~\cite{Muck:2001yv}. The coefficients $\delta_{k,l,n}$ and $\tilde{\delta}_{k,l,n}$ are simple combinations of Kronecker deltas, cf. (\ref{DeltaThree}), (\ref{DeltaFour}) and (\ref{DeltaLimit}), which define selection rules for couplings in theories without BKT. With the help of the effective transformations (\ref{gaugeKK4D}), we could in principle follow the technique discussed in last section and derive the Ward identities of the 4D theory. This derivation is complicated by the fact that we now have to deal with an infinite set of coupled transformations instead of simply (\ref{gauge4D}).

Instead of restricting our discussion to the 4D theory, let us define what we mean by functional differentiation on $S^1/\Zb_2$.

\begin{equation}
\begin{split}
\label{5Ddiff}
\frac{\delta A^a_{\mu}(x_1,y_1)}{\delta A^b_{\nu}(x_2,y_2)} &\equiv \frac{g_\mu^\nu \, \delta^{ab}}{2} \, \big[\delta(y_1-y_2; \, r_c) + \delta(y_1 + y_2; \, r_c)\big] \, \delta^{(4)}(x_1-x_2)\\
\frac{\delta A^a_5(x_1,y_1)}{\delta A^b_5(x_2,y_2)} &\equiv \hspace{5mm} \frac{\delta^{ab}}{2} \, \big[\delta(y_1-y_2; r_c) - \delta(y_1+y_2; \, r_c)\big] \, \delta^{(4)}(x_1-x_2)
\end{split}
\end{equation}

\noindent
Notice that both LHS, by definition (\ref{parities}), and RHS of (\ref{5Ddiff}) are even/odd in $y_1$ and $y_2$. The delta function $\delta(y; \, r_c)$ depends explicitly on the expansion on $S^1/\Zb_2$, and is given by the completeness of our set of orthonormal functions (\ref{completeness}). Now we can simply repeat our arguments of the previous section. The invariance of the classical part of Lagrangian $\Lag_{\rm 5D\YM}$ under gauge transformations (\ref{gauge5D}) implies the invariance of the tree-level effective action $\Gamma[A^a_M] = - 1/4 \int d^4x \; \int dy \; [1 + r_c \delta(y)] F^a_{MN} F^{a \; MN}$, which in turn gives rise to a master Ward identity for 5D Yang-Mills theories.

\begin{equation}
\partial_M \, \frac{\delta \Gamma}{\delta A^a_M} - g_5 f^{abc} \frac{\delta \Gamma}{\delta A^b_M} A^c_M = 0
\end{equation}

\noindent
Using the decomposition of the components of the functional derivative $\delta /\delta A^a_M$ in terms of the orthonormal functions $f_n$ and $g_n$,

\begin{equation}
\begin{split}
\frac{\delta}{\delta A^a_{\mu}(x,y)} &= \sum_{n=0}^{\infty} f_n(y) \; \frac{\delta}{\delta A^a_{(n) \mu}(x)}\\
\frac{\delta}{\delta A^a_5 (x,y)} &= \sum_{n=1}^{\infty} g_n(y) \; \frac{\delta}{\delta A^a_{(n) 5}(x)} \, ,
\end{split}
\end{equation}

\noindent
and finally integrating over the compact dimension, we derive a master Ward identity for the effective 4D theory.

\begin{equation}
\begin{split}
\partial_{\mu} \frac{\delta \Gamma}{\delta A^a_{(n) \mu}}\ +\ m_n
\frac{\delta \Gamma}{\delta A^a_{(n) 5}}\ &=\ \; g f^{abc} \,
\sum_{m,l=0}^{\infty} \sqrt{2}^{\; -1-\delta_{n,0} - \delta_{m,0} -
\delta_{l,0}}\\ &\times\, \bigg(\, \frac{\delta \Gamma}{\delta A^b_{(m)
\mu}} A^c_{(l) \mu} \Delta_{m, n, l}\ +\ \frac{\delta \Gamma}{\delta
A^b_{(m) 5}} A^c_{(l) 5} \tilde{\Delta}_{m,n,l}\, \bigg)
\end{split}
\end{equation}

\noindent
Standard functional differentiation with respect to the KK modes, cf. detailed discussion below (\ref{masterWard4DYM}), yields five distinct Ward identities that relate the fundamental interactions of the theory. Gauge and scalar KK modes will be represented by wavy and dashed lines respectively, see Feynman rules in Appendix~\ref{FeynmanRules}. The last two of the identities below involve an infinite summation and can be explicitly checked using the relations derived in Appendix~\ref{summation}.
\begin{equation}
\label{Ward1}
\begin{split}
-i &k_{\mu} \BBB{a \; \mu}{b \; \nu}{c \; \rho}{(n) \; k}{(m) \; p}{(l)
\; q} = \ \  -m_n \SBB{a}{b \; \nu}{c \; \rho}{(n) \; k}{(m) \; p}{(l)
\; q} \\ &\phantom{} \\ &\phantom{} \\
&+\ \sqrt{2}^{\; -1-\delta_{n,0}
- \delta_{m,0} - \delta_{l,0}} \, \Delta_{n, m, l} \: g \bigg[ f^{abd} \nshift \BB{d \; \nu}{c \; \rho}{(l) \; q} \nshift +\ f^{acd} \nshift \BB{d \; \rho}{b \; \nu}{(m) \; p} \nshift \; \bigg]
\end{split}
\end{equation}

\begin{equation}
\begin{split}
&-i k^{\mu} \BBS{a \; \mu}{b \; \nu}{c}{(n) \; k}{(m) \; p}{(l) \; q} =\
-m_n \SBS{a}{b \; \nu}{c}{(n) \; k}{(m) \; p}{(l) \; q}\\ &\phantom{}
\\ &\phantom{}\\
&+\ \sqrt{2}^{\; -1-\delta_{n,0} - \delta_{m,0}} \, g
\, \bigg[ \Delta_{m,n,l} f^{abd} \nshift \BS{d \; \nu}{c}{(l) \; q} \nshift -\
\tilde{\Delta}_{m,n,l} f^{acd} \nshift \BS{b \; \nu}{d}{(m) \; p} \nshift \bigg]
\end{split}
\end{equation}
\begin{equation}
\begin{split}
-i &k_{\mu} \BSS{a \; \mu}{b}{c}{(n) \; k}{(m) \; p}{(l) \; q} =\\ &\phantom{}
\\ &\phantom{}\\
 &\sqrt{2}^{\; -1-\delta_{n,0}} \, \tilde{\Delta}_{m,n,l} \: g \, \bigg[
f^{abd} \SSLL{ d }{ c }{(l) \; q} +\ f^{acd} \SSLL{d}{b}{(m) \; p} \bigg]\\
&\phantom{}
\end{split}
\end{equation}

\medskip
\begin{equation}
\begin{split}
-i k^{\mu} \BBBB{a \; \mu}{b \; \nu}{c \; \rho}{d \; \sigma}{(n) \;
k}{(m) \; p}{(l) \; q}{(k) \; r} \nshift =\sum_{j=0}^{\infty}
g \, &\sqrt{2}^{\; -1-\delta_{n,0} - \delta_{j,0}} \bigg[ \sqrt{2}^{\; -
\delta_{m,0}} \, \Delta_{m,n,j} f^{abe}\\ &\phantom{}\\
\times \BBB{e \; \nu}{c \; \rho}{d \; \sigma}{(j) \; k+p}{(l) \; q}{(k) \; r} &+ \sqrt{2}^{\; - \delta_{l,0}} \, \Delta_{l,n,j} f^{ace} \BBB{e \;
\rho}{b \; \nu}{d \; \sigma}{(j) \; k+q}{(m) \; p}{(k) \; r}\\ &\phantom{} \\
 +\ \sqrt{2}^{\; - \delta_{k,0}} \, \Delta_{k,n,j} &f^{ade}
\BBB{e \; \sigma}{b \; \nu}{c \; \rho}{(j) \; k+r}{(m) \; p}{(l) \; q} \bigg]\\ &\phantom{} \\ &\phantom{}
\end{split}
\end{equation}
\begin{equation}
\label{Ward5}
\begin{split}
-i k^{\mu} \BBSS{a \; \mu}{b \; \nu}{c}{d}{(n) \; k}{(m) \; p}{(l) \; q}{(k) \; r} \hspace{-3mm} = \sum_{j=0}^{\infty} g \, &\sqrt{2}^{\; -1-\delta_{n,0} - \delta_{j,0}} \bigg[ \sqrt{2}^{\; - \delta_{m,0}} \, \Delta_{m,n,j} f^{abe}\\ &\phantom{}\\
\times \BSS{e \; \nu}{c}{d}{(j) \; k+p}{(l) \; q}{(k) \; r} &+ \tilde{\Delta}_{l,n,j} f^{ace} \SBS{e}{b \; \nu}{d}{(j) \; k+q}{(m) \; p}{(k) \; r}\\ &\phantom{}\\
 + \tilde{\Delta}_{k,n,j} &f^{ade} \SBS{e}{b \; \nu}{c}{(j) \; k+r}{(m) \; p}{(l) \; q} \bigg]\\
&\phantom{}\\ &\phantom{}
\end{split}
\end{equation}
\noindent
Let us now turn to the quantization of our theory and discuss the last two terms appearing in Lagrangian (\ref{Lag5DYM}). We have got two options: We can either fix our gauge before or after compactification. In the second case we would have to add the two terms $\Lag_{\rm 4D \GF}(x)$ and $\Lag_{\rm 4D \FP}(x)$ to the 4D effective Lagrangian that we get after integration over the compact dimension. Since we would like to see all vector-scalar mixing terms cancelled, cf. (\ref{mixing}), we expect each of the above terms to consist of an infinite number of contributions.

Instead we opt to approach the problem from the five-dimensional point of view; a strategy that paid off in the derivation of the Ward identities. As already adopted in~\cite{Muck:2001yv}, we will work in the framework of generalized $R_\xi$ gauges\footnote{The naive generalization of the $R_\xi$ gauges to five dimensions, $F[A^a_M] = \partial^\mu A^a_\mu - \partial_5 A^a_5$, is modified in order to ensure the cancellation of the vector-scalar mixing. The analogous gauge is widely used in spontaneously broken gauge theories~\cite{Delbourgo:1987np}, where would-be Goldstone bosons play the role of the scalar KK modes in our theory.} and choose the following gauge-fixing functional.

\begin{equation}
F[A^a_M] = \partial^\mu A^a_\mu - \xi \; \partial_5 A^a_5
\end{equation}

\noindent
We take care of the localized terms in our Lagrangian by multiplying the gauge-fixing and Faddeev-Popov terms by a factor $[1 + r_c \delta(y)]$. It is the same factor that appears in the classical part of Lagrangian (\ref{Lag5DYM}).

\begin{align}
\label{5DGF}
\begin{split}
\Lag_{\rm 5D\GF}(x,y) &= -\, \big[1 + r_c \delta(y) \big] \, \frac{1}{2 \xi} \; \big( F[A^a_M]\big)^2\\
&= -\, \big[ 1 + r_c \delta(y) \big] \, \frac{1}{2 \xi}\,  \Big( \partial^{\mu} A^a_{\mu} - \xi \partial_5 A^a_5 \Big)^2
\end{split}\\[3mm]
\label{5DFP}
\begin{split} 
\Lag_{\rm 5D\FP}(x,y) &= \, \big[ 1 + r_c \delta(y) \big] \, \bar{c}^a \; \frac{\delta F[A^a_M]}{\delta \theta^b} \; c^b\\
&= \, \big[ 1 + r_c \delta(y) \big] \bar{c}^a \Big[\delta^{ab} ( \partial^2 - \xi \partial_5^2 ) - g_5 f^{abc} ( \partial^{\mu} A^c_{\mu} - \xi \partial_5 A^c_5 ) \Big] c^b
\end{split}
\end{align}

\noindent
It is straightforward to check that any mixing between gauge and scalar sector or between different KK modes is absent in the quantized Lagrangian. After compactification, (\ref{5DGF}) and (\ref{5DFP}) lead to an effective 4D theory in which each of the KK modes is quantized in a conventional $R_\xi$ gauge.

Due to the common factor, the complete Lagrangian (\ref{Lag5DYM}) is invariant under standard 5D BRS transformations.

\begin{equation}
\label{BRS5DYM}
\begin{split}
s\, A^a_M &= D^{ab}_M \, c^b\\
s\, c^a  &= - \frac{g_5 f^{abc}}{2} \; c^b c^c\\
s\, \overline{c}^a &= \frac{F[A^a_M]}{\xi}
\end{split}
\end{equation}

\noindent
The ghosts $c^a(x,y)$ and $\overline{c}^a(x,y)$ are even fields in $y$, which follows from the consistency of the above transformations. After integration over the compact dimension, we find the effective BRS transformations of the KK modes.

\begin{equation}
\label{BRSeffective}
\begin{split}
s A^a_{(n) \mu} &= \partial_{\mu} c^a_{(n)}\: -\: g f^{abc} \sum_{m,l=0}^{\infty} \sqrt{2}^{\; -1-\delta_{n,0}-\delta_{m,0}-\delta_{l,0}}\, \Delta_{n,m,l} \; A^c_{(m) \mu} c^b_{(l)}\\
s A^a_{(n) 5} &= - m_n c^a_{(n)}\: -\: g f^{abc} \sum_{m, l=0}^{\infty} \sqrt{2}^{\; -1-\delta_{l,0}}\, \tilde{\Delta}_{n,m,l} \; A^c_{(m) 5} c^b_{(l)}\\
s c^a_{(n)} &= -\; \frac{g f^{abc}}{2} \sum_{m, l=0}^{\infty} \sqrt{2}^{\; -1-\delta_{n,0}-\delta_{m,0}-\delta_{l,0}}\, \Delta_{n,m,l} \; c^b_{(m)} c^c_{(l)}\\
s \overline{c}^a_{(n)} &= \frac{1}{\xi}\, \partial^{\mu} A^a_{(n) \mu}\: -\: m_n A^a_{(n) 5}
\end{split}
\end{equation}

\noindent
After successful quantization, we proceed with our program and derive the ST identities of the effective 4D theory. We will follow closely our discussion of the previous section. The definition of the generating functionals generalizes in an obvious way.

\begin{equation}
\label{defZ}
\begin{split}
W = e^{iZ} &= \int DA \, Dc \, D\overline{c} \; \exp \bigg[ \, i \int d^4x \, \int_{-\pi R}^{\pi R} dy \; \bigg( \Lag_{\rm 5D \YM}\\
&+ [1 + r_c \delta (y)] [J^{a M} A^a_M + \overline{D}^a c^a + \overline{c}^a D^a + K^{a M} s A^a_M + M^a sc^a ] \bigg) \bigg]
\end{split}
\end{equation}

\noindent
The response of the generating functional $Z$ to infinitesimal BRS transformations (\ref{BRS5DYM}) gives rise to a master ST identity in five dimensions, which in turn translates into a master equation for the connected Green's functions of the KK modes.

\begin{equation}
J^{a \, \mu} \frac{\delta Z}{\delta K^{a \, \mu}} - J^{a \, 5} \frac{\delta Z}{\delta K^{a \, 5}} - \overline{D}^a \frac{\delta Z}{\delta M^a} + \frac{1}{\xi} D^a \partial^{\mu} \frac{\delta Z}{\delta J^{a \mu}} - D^a \partial^5 \frac{\delta Z}{\delta J^{a \, 5}} =0
\end{equation}

\begin{equation}
\label{STmaster5D}
\begin{split}
\sum_{n=0}^{\infty} \; \bigg[ J^{a \, \mu}_{(n)} \, \frac{\delta Z}{\delta K^{a \, \mu}_{(n)}} &+ J^{a \, 5}_{(n)} \, \frac{\delta Z}{\delta K^{a \, 5}_{(n)}} - \overline{D}^a_{(n)} \, \frac{\delta Z}{\delta M^a_{(n)}}\\
&+ \frac{1}{\xi} D^a_{(n)} \, \partial^{\mu} \frac{\delta Z}{\delta J^{a \mu}_{(n)}} - m_n D^a_{(n)} \, \frac{\delta Z}{\delta J^{a \, 5}_{(n)}} \bigg] = 0
\end{split}
\end{equation}

\noindent
After defining the 5D generating functional $\Gamma$ and integrating over the extra dimension, we derive a master ST identity for 1PI Green's functions in the effective 4D theory.

\begin{equation}
\label{defGamma}
\begin{split}
\Gamma[A^a_M, &c^a, \bar{c}^a, K^{a M}, M^a] = Z[J^{a M}, \overline{D}^a, D^a, K^{a M}, M^a]\\
&- \int d^4 x \int dy \big[1 + r_c \delta(y) \big] \big( J^{a M} A^a_M + \overline{D}^a c^a + \overline{c}^a D^a \big)
\end{split}
\end{equation}

\begin{equation}
\label{STmaster5D1PI}
\begin{split}
\sum_{n=0}^{\infty} \; \bigg[ \frac{\delta \Gamma}{\delta A^a_{(n) \mu}} \frac{\delta \Gamma}{\delta K^{a \mu}}_{(n)} &+ \frac{\delta \Gamma}{\delta A^a_{(n) 5}} \frac{\delta \Gamma}{\delta K^{a \, 5}_{(n)}} + \frac{\delta \Gamma}{\delta c^a_{(n)}} \frac{\delta \Gamma}{\delta M^a_{(n)}}\\
&- \frac{1}{\xi} \frac{\delta \Gamma}{\delta \overline{c}^a_{(n)}} \partial^{\mu} A^a_{(n) \mu} + m_n \frac{\delta \Gamma}{\delta \overline{c}^a_{(n)}} A^a_{(n) 5} \bigg] = 0
\end{split}
\end{equation}

\noindent
Functional differentiation of (\ref{STmaster5D}) and (\ref{STmaster5D1PI}) will generate specific ST identities. But again, it proves to be more practical to start off from the BRS invariance of the Green's functions themselves. Consider the following example.

\begin{equation}
s \, \langle 0 | T \bar{c}^a_{(n)}(x) \, A^b_{(n) \nu} (z) \, A^c_{(n) \rho} (u) \, A^d_{(n) \sigma} (v) |0 \rangle = 0
\end{equation}

\newpage
\noindent
The action of the BRS operator $s$ on the KK modes is given by (\ref{BRSeffective}). Remember that terms bilinear in fields do not contribute to \textit{on-shell} ST identities. The derivation is therefore simple indeed, and the final ST identity can be graphically represented as below. We will omit KK numbers and momenta on the RHS, since they are identical in all graphs. Dotted lines will stand for KK ghosts of the theory.

\begin{equation}
\label{ST1}
\begin{split}
\frac{p_1^{\mu}}{m_n} \hspace{-1mm} \BBBBblob{\mu}{\nu}{\rho}{\sigma}{(n) \; p_1}{(n) \; p_2}{(n) \; k_1}{(n) \; k_2} &= \; i \hspace{-7mm} \SBBBblob{}{\nu}{\rho}{\sigma}{}{}{}{} \hspace{-7mm} + \frac{p_2^{\nu}}{m_n} \hspace{-5mm} \GGBBblob{}{}{\rho}{\sigma}{}{}{}{}\\
&\phantom{}\\ & \qquad \qquad + \frac{k_2^{\sigma}}{m_n} \hspace{-5mm} \GBBGblob{}{\nu}{\rho}{}{}{}{}{} \hspace{-7mm} + \frac{k_1^{\rho}}{m_n} \hspace{-7mm} \GBGBblob{}{\nu}{}{\sigma}{}{}{}{} \hspace{-7mm} \\
&\phantom{}
\end{split}
\end{equation}
\vspace{4mm}

\noindent
In the next chapter, we are going to make use of this identity in our proof of the GET. For the particular example that we would like to discuss, we will need one further on-shell ST identity. It originates from the BRS invariance of the following Green's function.

\begin{equation}
s \, \langle 0 | T \bar{c}^a_{(n)}(x) \, A^b_{(n) 5} (z) \, A^c_{(n) \rho} (u) \, A^d_{(n) \sigma} (v) |0 \rangle = 0
\end{equation}

\begin{equation}
\label{ST2}
\begin{split}
\frac{p_1^{\mu}}{m_n} \hspace{-1mm} \BSBBblob{\mu}{}{\rho}{\sigma}{(n) \; p_1}{(n) \; p_2}{(n) \; k_1}{(n) \; k_2} &= \; i \hspace{-7mm} \SSBBblob{}{}{\rho}{\sigma}{}{}{}{} \hspace{-7mm} -i \hspace{-4mm} \GGBBblob{}{}{\rho}{\sigma}{}{}{}{}\\ &\phantom{}\\
& \qquad \qquad + \frac{k_2^{\sigma}}{m_n} \hspace{-5mm} \GSBGblob{}{}{\rho}{}{}{}{}{} \hspace{-7mm} + \frac{k_1^{\rho}}{m_n} \hspace{-7mm} \GSGBblob{}{}{}{\sigma}{}{}{}{} \hspace{-7mm} \\ &\phantom{}
\end{split}
\end{equation}
\vspace{4mm}

\newpage
\noindent
Note that ST identities of this kind are familiar from spontaneously broken gauge theories in four dimensions. In these theories would-be Goldstone bosons replace our scalar modes $A^a_{(n) 5}$. The similarities between compactified 5D gauge theories and spontaneously broken 4D theories go even further, as we will see in the next chapter.


\chapter{Generalized Equivalence Theorem}
\label{GET}

In the high energy limit of a spontaneously broken gauge theory, a scattering amplitude with longitudinally polarized massive vector bosons in the initial/final state is up to a phase equal to an amplitude in which the bosons are replaced by the corresponding unphysical would-be Goldstone modes. This relation is famously known as equivalence theorem (ET)~\cite{Lee:1977eg, Cornwall:1974km} and is an important tool for the perturbative calculation of high energy scattering processes. 

When discussing quantization schemes and ST identities in the last chapter, we already noticed the parallels between spontaneously broken theories and compactified orbifold theories. As will show here, the ET does also hold for orbifold theories with BKT\footnote{In~\cite{Chivukula:2001hz} the ET was checked for particular processes in orbifold theories without BKT.}. To be explicit, the ET states that

\begin{equation}
\label{ET}
\begin{split}
T&\big[A^{a_1}_{(n_1) L}, \, \ldots, \, A^{a_k}_{(n_k) L}, \, S \to 
  A^{b_1}_{(m_1) L}, \, \ldots, \, A^{b_l}_{(m_l) L}, \, S'\big] =\\
&C \, i^k \, (-i)^l \; T\big[A^{a_1}_{(n_1) 5}, \, \ldots, \,  A^{a_k}_{(n_k) 5}, \, S
\to A^{b_1}_{(m_1) 5}, \, \ldots, \, A^{b_l}_{(m_l) 5}, \, S'\big] + \ord \big( \frac{m_{n}}{E} \big) \; ,
\end{split}
\end{equation}

\noindent
where $A^{a_i}_{(n_i) \, L}$ are longitudinally polarized KK vector bosons, $A^{a_i}_{(n_i) \, 5}$ their associated scalar modes and $S, S'$ any spectators in the initial/final state that do not include longitudinal polarizations. $m_n$ is a typical mass of the scattering particles and $E \approx \sqrt{s}$ the centre of mass energy of the scattering process. In addition to the aforementioned phase, $i^k \, (-i)^l$, there appears a factor $C$, which in general is renormalization scheme dependant~\cite{Yao:1988aj, Yao:1988aj, He:1992ng}. However, $C=1$ at tree-level and in certain renormalization schemes that respect the Ward identities of the classical action~\cite{Papavassiliou:1997pb, Denner:1996gb}.

The ET makes only statements about the order of the scattering amplitudes. In fact, we can be far more precise and specify the energetically suppressed terms in the above relation. This complete relation extends the ET and is known as the generalized equivalence theorem (GET)~\cite{Chanowitz:1985hj, Gounaris:1986cr}. In order to describe the GET, we will need one further fact: At high energies a longitudinal polarization vector $\epsilon^\mu_L$ is dominated by a component parallel to its momentum. The remainder $a^{\mu}(k) \sim \ord(m_n / E)$ is energetically suppressed.

\begin{equation}
\label{eps}
\epsilon^{\mu}_L \, (k) = \frac{k^{\mu}}{m_n} + a^{\mu}(k)
\end{equation}

\noindent
Consider a scattering process with a certain number of longitudinal vector bosons in the initial/final state. The following set of rules will generate the GET, i.e. the complete RHS of (\ref{ET}):

\begin{itemize}
\item[(i)] Write down the sum of all amplitudes that result from replacing \textit{any} number of longitudinal vector bosons by the respective KK scalar modes (would-be Goldstone bosons).
\item[(ii)] In most of the amplitudes some longitudinal bosons will remain. Replace their polarization vector $\epsilon^\mu_L$ by the remainder $a^\mu$.
\item[(iii)] Multiply each of the amplitudes by a phase $i^k \, (-i)^l$, where $k$ and $l$ are the number of initial (final) KK scalar modes (would-be Goldstone bosons) of the particular amplitude.
\end{itemize}

\noindent
As an example, let us consider the scattering of two transverse KK gauge bosons into two longitudinal KK gauge bosons. The GET for the process reads

\begin{equation}
\label{GETexample}
\begin{split}
T[A^a_{(n) T} &A^b_{(n) T} \rightarrow A^c_{(n) L} A^d_{(n) L}] =\\
- &T[A^a_{(n) T} A^b_{(n) T} \rightarrow A^c_{(n) 5} A^d_{(n) 5}] - i T[A^a_{(n) T} A^b_{(n) T} \rightarrow A^c_{(n) 5} a^d_{(n)}]\\
-i &T[A^a_{(n) T} A^b_{(n) T} \rightarrow a^c_{(n)} A^d_{(n) 5}] + T[A^a_{(n) T} A^b_{(n) T} \rightarrow a^c_{(n)} a^d_{(n)}] \; .
\end{split}
\end{equation}

\noindent
In the above amplitudes, $a^b_{(n)}$ stands for a vector boson $A^b_{(n) \mu}$ whose polarization vector $\epsilon^\mu_L$ was replaced by $a^\mu$. Since each of the remainders $a^\mu$ is energetically suppressed, the ET follows:

\begin{equation}
\begin{split}
T[A^a_{(n) T} A^b_{(n) T} \rightarrow A^c_{(n) L} A^d_{(n) L}] = - T[A^a_{(n) T} A^b_{(n) T} \rightarrow A^c_{(n) 5} A^d_{(n) 5}] \, + \, \ord\big( \frac{m_n}{E} \big) \; .
\end{split}
\end{equation}

\noindent
The proof of the GET (\ref{GETexample}) relies on the two on-shell ST identities (\ref{ST1}) and (\ref{ST2}). Let $\epsilon_{a, \, b}$ and $\epsilon_{1, \, 2}$ be the transverse and longitudinal polarization vectors respectively. In a first step, let us rewrite $\epsilon_1=- k_1/m_n + a_1$ and apply (\ref{ST1}) to the first term of the amplitude. Note that $k_1$ flows into the vertex. In the notation of this chapter, labels assigned to the endpoints of graphs do not stand for Lorentz indices but for four-vectors the amplitudes are contracted with.

\newpage
\begin{equation}
\label{proof0}
\BBBBblob{\epsilon_a}{\epsilon_b}{\epsilon_1}{\epsilon_2}{(n)p_1}{(n)p_2}{(n)k_1}{(n)k_2} \hspace{-2mm} = \; -i \hspace{-4mm} \BBSBblob{\epsilon_a}{\epsilon_b}{}{\epsilon_2}{}{}{}{} \hspace{-4mm} + \hspace{-4mm} \BBBBblob{\epsilon_a}{\epsilon_b}{a_1}{\epsilon_2}{}{}{}{}
\end{equation}
\vspace{15mm}

\noindent
Note that amplitudes with external ghosts are absent, since $\epsilon_2 \cdot k_2 = \epsilon_a \cdot p_1 = \epsilon_b \cdot p_2 = 0$. In the following step, we split the second polarization vector $\epsilon_2 = - k_2/m_n + a_2$.
 
\begin{equation}
\begin{split}
\label{proof1}
\cdots \; &= \; i \hspace{-4mm} \BBSBblob{\epsilon_a}{\epsilon_b}{}{k_2 / m_n}{}{}{}{} -i \BBSBblob{\epsilon_a}{\epsilon_b}{}{a_2}{}{}{}{}\\ &\phantom{}\\ &\phantom{}\\
&\hspace{2.5cm} - \hspace{-4mm} \BBBBblob{\epsilon_a}{\epsilon_b}{a_1}{k_2 / m_n}{}{}{}{} + \BBBBblob{\epsilon_a}{\epsilon_b}{a_1}{a_2}{}{}{}{}
\end{split}
\end{equation}
\vspace{15mm}

\noindent
After applying (\ref{ST1}) and (\ref{ST2}) to the third and first term respectively, two contributions with external ghosts remain.

\begin{equation}
\begin{split}
\label{proof2}
\cdots \; = \; - \hspace{-4mm} &\BBSSblob{\epsilon_a}{\epsilon_b}{}{}{}{}{}{} \hspace{-4mm} + \hspace{-4mm} \BBGGblob{\epsilon_a}{\epsilon_b}{}{}{}{}{}{} \hspace{-4mm} -i \hspace{-4mm} \BBSBblob{\epsilon_a}{\epsilon_b}{}{a_2}{}{}{}{}\\ &\phantom{}\\ &\phantom{}\\
&\quad \quad -i \hspace{-4mm} \BBBSblob{\epsilon_a}{\epsilon_b}{a_1}{}{}{}{}{} \hspace{-4mm} - \frac{a_1 \cdot k_1}{m_n} \hspace{-4mm} \BBGGblob{\epsilon_a}{\epsilon_b}{}{}{}{}{}{} \hspace{-4mm} + \hspace{-4mm} \BBBBblob{\epsilon_a}{\epsilon_b}{a_1}{a_2}{}{}{}{}
\end{split}
\end{equation}

\noindent
Since $a_1 \cdot k_1 = m_n$, the second and the fifth term cancel. What we are left with is the RHS of (\ref{GETexample}) in diagrammatic form.

\begin{equation}
\cdots \; = \; - \hspace{-4mm} \BBSSblob{\epsilon_a}{\epsilon_b}{}{}{}{}{}{} \hspace{-4mm} -i \hspace{-4mm} \BBSBblob{\epsilon_a}{\epsilon_b}{}{a_2}{}{}{}{} \hspace{-4mm} -i \hspace{-4mm} \BBBSblob{\epsilon_a}{\epsilon_b}{a_1}{}{}{}{}{} \hspace{-4mm} + \hspace{-4mm} \BBBBblob{\epsilon_a}{\epsilon_b}{a_1}{a_2}{}{}{}{} \nonumber
\end{equation}
\vspace{15mm}

\noindent
Using the same arguments and an extended set of ST identities, one can prove the GET for any other process. We can now understand the structure of the GET: Amplitudes without any external scalars result merely from the splitting of the polarization vectors and remain unchanged, cf. the last terms in (\ref{proof0}) and (\ref{proof1}). Any external scalar results from the application of an ST identity and the amplitudes aquire a factor $i$, cf. first terms in (\ref{ST1}) and (\ref{ST2}). In the case of final state scalars, the momentum in (\ref{eps}) is understood to flow out of the vertices and the amplitudes aquire therefore an additional sign. The phases mentioned in point (iii) at the beginning of this chapter are now evident.

Checking the GET order by order is of course possible, but even a tree-level calculation using the Ward identities (\ref{Ward1}) to (\ref{Ward5}) turns out to be tedious indeed. A tree-level check of the ET on the other hand is feasible and will be instructive. Unlike in orbifold theories without localized terms, the KK number in the fundamental interactions of BKT orbifold theories is not conserved. For that reason, already at lowest order there is an infinite number of diagrams contributing to any given process. Only subtle cancellations among them ensure the ET to hold. Let us have a look at a specific example, the elastic scattering of two longitudinally polarized gauge bosons, $A^a_{(n)L} \, A^b_{(n)L} \, \rightarrow \, A^c_{(n)L} \, A^d_{(n)L}$, for which the ET\footnote{The complete GET involves 16 terms, instead of the four terms of example (\ref{GETexample}).} reads as follows.

\begin{equation}
\label{ETexample}
\BBBBblob{\epsilon_a}{\epsilon_b}{\epsilon_c}{\epsilon_d}{(n)p_1}{(n)p_2}{(n)k_1}{(n)k_2} \hspace{-2mm} = \hspace{-2mm} \SSSSblob{}{}{}{}{(n)p_1}{(n)p_2}{(n)k_1}{(n)k_2} \hspace{-2mm} \; + \; \ord \big( \frac{m_n^2}{s} \big)
\end{equation}
\vspace{10mm}

\noindent
The kinematics is particular simple, since all external particles are of the same mass $m_n$. The cosine of the centre of mass system scattering angle will be denoted by $c = \cos \theta$.

\begin{equation}
\label{kinematics}
\begin{split}
s &= (p_1+p_2)^2\\
t &= (p_1+k_1)^2=-(1-c) \frac{\beta^2 s}{2}\\
u &= (p_1+k_2)^2=-(1+c) \frac{\beta^2 s}{2}
\end{split}
\end{equation}

\begin{equation}
\begin{split}
\beta^2 &= 1-\frac{4 m_n^2}{s}\\
\epsilon_a^{\mu} &= \frac{1}{2 m_n \beta} \left[ (1+\beta^2)p_1^{\mu} - (1-\beta^2) p_2^{\mu} \right]
\end{split}
\end{equation}

\noindent
The tree-level amplitude for the gauge boson scattering is given by an infinite set of Feynman diagrams. Each of them is straightforward to calculate.

\begin{align}
\BBBBblob{\epsilon_a}{\epsilon_b}{\epsilon_c}{\epsilon_d}{(n)p_1}{(n)p_2}{(n)k_1}{(n)k_2} \hspace{-2mm} &=  \hspace{-2mm} \BBBB{\epsilon_a}{\epsilon_b}{\epsilon_c}{\epsilon_d}{(n)p_1}{(n)p_2}{(n)k_1}{(n)k_2} + \; \sum_{j=0}^{\infty} \sChannelB{\epsilon_a}{\epsilon_b}{\epsilon_c}{\epsilon_d}{(n)p_1}{(n)p_2}{(n)k_1}{(n)k_2}{\, \, \, \, (j)} + \; \cdots \nonumber\\ &\phantom{}\nonumber\\ &\phantom{}\nonumber\\
\begin{split}
\label{gaugeScattering}
&= i T_4+\sum_{j=0}^{\infty} \; [i T^s_{(j)}+i T^t_{(j)}+i T^u_{(j)}]\\
&= i T_4+i T^s+i T^t+i T^u
\end{split}
\end{align}

\begin{align}
i T_4 &= \Delta_{n,n,n,n} \, \frac{i g^2}{8 m_n^4} \Big[ f^{abe} f^{cde} s(t-u)+ f^{ace} f^{bde} t\big(s-\frac{u}{\beta^4}\big) + f^{ade} f^{bce} u\big(s-\frac{t}{\beta^4}\big) \Big]\nonumber \\
\begin{split}
\label{jcontributions}
i T^s_{(j)} &= 2^{-\delta_{j,0}} \Delta^2_{n,n,j} \, i g^2 f^{abe} f^{cde} \frac{s(u-t)}{8 m_n^4} \Big[ 1+ \frac{2m_n^2}{s} \Big]^2 \, \frac{s}{s-m_j^2}\\
i T^t_{(j)} &= 2^{-\delta_{j,0}} \Delta^2_{n,n,j} \, i g^2 f^{ace} f^{bde} \Big[ \frac{u-s}{2t} \Big(1+ \frac{t}{2 m_n^2 \beta^2} \Big)^2 + \frac{t-2u}{m_n^2 \beta^2} \Big] \frac{t}{t-m_j^2}\\
i T^u_{(j)} &= 2^{-\delta_{j,0}} \Delta^2_{n,n,j} \, i g^2 f^{ade} f^{bce} \Big[ \frac{t-s}{2u} \Big(1+ \frac{u}{2 m_n^2 \beta^2} \Big)^2 + \frac{u-2t}{m_n^2 \beta^2} \Big] \frac{u}{u-m_j^2}
\end{split}
\end{align}

\noindent
We are not interested in a complete expression for (\ref{gaugeScattering}), but do only need the leading order for the ET. Our calculation is much simplified if we expand factors that appear in (\ref{jcontributions}) up to order $m_j^4/s^2$,

\begin{equation}
\label{threePlusRest}
\begin{split}
\frac{s}{s-m_j^2} &= 1 + \frac{m_j^2}{s} + \frac{m_j^4}{s^2} + \ord \big( \frac{m_j^6}{s^3} \big)\\
\frac{t}{t-m_j^2} &= 1 - \big( \frac{2}{1-c} \, \frac{1}{s} + \frac{8 m_n^2}{1-c} \, \frac{1}{s^2} \big) m_j^2 + \frac{4}{(1-c)^2} \, \frac{1}{s^2} \, m_j^4 + \ord \big( \frac{m_j^6}{s^3} \big)\\
\frac{u}{u-m_j^2} &= 1 - \big(\frac{2}{1+c} \, \frac{1}{s} + \frac{8m_n^2}{1+c} \, \frac{1}{s^2} \big) m_j^2 + \frac{4}{(1+c)^2} \, \frac{1}{s^2} \, m_j^4 + \ord \big( \frac{m_j^6}{s^3} \big) \; ,
\end{split}
\end{equation}
 
\noindent
and split each of the contributions accordingly,
 
\begin{equation}
T^{s,t,u}_{(j)} = T^{s,t,u}_{(j) \I} + T^{s,t,u}_{(j) \II} + T^{s,t,u}_{(j) \III} + \ord \big( \frac{m_j^2}{s} \big) \; .
\end{equation}

\noindent
The masses $m_j$ now appear as simple multiplicative factors, and we can use relations derived in Appendix~\ref{summation} in order to sum over the infinite KK tower.

\begin{equation}
\begin{split}
\label{T4}
i T_4 = \Delta_{n,n,n,n} i g^2 \Big[ &f^{abe} f^{cde} \big( \frac{c}{8 m_n^4} s^2 - \frac{c}{2 m_n^2} s \big)+\\
&f^{ace} f^{bde} \big(\frac{(c+3)(c-1)}{32m_n^4} s^2 + \frac{1-c}{4m_n^2} s \big) +\\
&f^{ade} f^{bce} \big(\frac{(c-3)(c+1)}{32m_n^4} s^2 + \frac{c+1}{4m_n^2} s \big) \Big]
\end{split}
\end{equation}

\begin{equation}
\begin{split}
\sum_{j=0}^{\infty} i T^s_{(j) \I} &= \Delta_{n,n,n,n} i g^2 f^{abe} f^{cde} \Big[ - \frac{c}{8m_n^4} s^2 + \frac{3c}{2} \Big] + \ord \big( \frac{m_n^2}{s} \big)\\
\sum_{j=1}^{\infty} i T^s_{(j) \II} &= \Delta_{n,n,n,n} i g^2 f^{abe} f^{cde} \Big[ - \frac{c}{6m_n^2} s \Big] + \ord \big( \frac{m_n^2}{s} \big)\\
\sum_{j=1}^{\infty} i T^s_{(j) \III} &= \big(\Delta_{n,n,n,n}+\frac{3}{4}X_n \big) i g^2 f^{abe} f^{cde} \big( -\frac{2c}{3} \big) + \ord \big( \frac{m_n^2}{s} \big)
\end{split}
\end{equation}

\begin{equation}
\label{tchannel}
\begin{split}
\sum_{j=0}^{\infty} i T^t_{(j) \I} &= \Delta_{n,n,n,n} \, i g^2 f^{ace} f^{bde}\\
&\qquad \times \Big[ \frac{(3+c)(1-c)}{32m_n^4} s^2 + \frac{c}{m_n^2} s + \frac{2c^2+c+1}{2(1-c)} \Big] + \ord \big( \frac{m_n^2}{s} \big)\\
\sum_{j=1}^{\infty} i T^t_{(j) \II} &= \Delta_{n,n,n,n} \, i g^2 f^{ace} f^{bde} \Big[ - \frac{3+c}{12 m_n^2} s+ \frac{c^2-6c-3}{3(1-c)} \Big] + \ord \big( \frac{m_n^2}{s} \big)\\
\sum_{j=1}^{\infty} i T^t_{(j) \III} &= \big(\Delta_{n,n,n,n} + \frac{3}{4}X_n \big) i g^2 f^{ace} f^{bde} \frac{2(3+c)}{3(1-c)} + \ord \big( \frac{m_n^2}{s} \big)
\end{split}
\end{equation}

\begin{equation}
X_n = 8\,N_n^4\, \pi^2 R^2 \tilde{r}_c^3 \, m_n^2
\end{equation}

\noindent
Replacing $c$ by $-c$ in (\ref{tchannel}), we find the u-channel sums $\sum_{j=0}^{\infty} i T^u_{(j) \I,\; \II,\; \III}$. Collecting all of the contributions, (\ref{T4}) to (\ref{tchannel}), we find $s^2$-contributions in s-, t- and u-channel to cancel against terms in $i T_4$. Terms linear in $s$ are identical for each colour factor and vanish due to the Jacobi identity. The leading contribution is therefore of $\ord(1)$.

\begin{equation}
\label{leadingOrderGauge}
\begin{split}
i T_4&+i T^s+i T^t+i T^u\\
&= (\Delta_{n,n,n,n} + X_n) i g^2 \Big[ f^{ace} f^{dbe} \frac{c^2+3}{2(c-1)} + f^{ade} f^{bce} \frac{c^2+3}{2(c+1)} \Big] + \ord \big( \frac{m_n^2}{s} \big)
\end{split}
\end{equation}

\noindent
Let us now check that the RHS of (\ref{ETexample}) agrees with this result and calculate the amplitude for the elastic scattering of two scalar KK modes.

\begin{align}
\SSSSblob{}{}{}{}{(n)p_1}{(n)p_2}{(n)k_1}{(n)k_2} \hspace{-2mm} \; &= \; \sum_{j=0}^{\infty} \sChannelS{}{}{}{}{(n)p_1}{(n)p_2}{(n)k_1}{(n)k_2}{\, \, \, \, (j)} \hspace{-2mm} \; + \; \cdots\nonumber \\ &\phantom{}\nonumber\\
\begin{split}
\label{scalarScattering}
&= \sum_{j=0}^{\infty} \; [i T^s_{(j)}+i T^t_{(j)}+i T^u_{(j)}]\\
&= i T^s+i T^t+i T^u
\end{split}
\end{align}

\noindent
In this case, no quartic coupling contributes. Using the Feynman rules of Appendix~\ref{FeynmanRules}, s-, t- and u-channel exchanges lead to the following contributions.

\begin{equation}
\label{jcontributions2}
\begin{split}
i T^s_{(j)} &= 2^{-\delta_{j,0}} \tilde{\Delta}^2_{n,j,n} \, i g^2 f^{abe} f^{cde} \frac{u-t}{2(s-m_j^2)}\\
i T^t_{(j)} &= 2^{-\delta_{j,0}} \tilde{\Delta}^2_{n,j,n} \, i g^2 f^{ace} f^{dbe} \frac{s-u}{2(t-m_j^2)}\\
i T^u_{(j)} &= 2^{-\delta_{j,0}} \tilde{\Delta}^2_{n,j,n} \, i g^2 f^{ade} f^{bce} \frac{t-s}{2(u-m_j^2)}
\end{split}
\end{equation}

\noindent
As done in (\ref{threePlusRest}), we expand factors in (\ref{jcontributions2}) that originate from the propagators. Using (\ref{Xsums}), we sum the leading orders and find $i T^s+i T^t+i T^u$ to be identical to (\ref{leadingOrderGauge}). The ET (\ref{ETexample}) is therefore verified. In the limit of vanishing $r_c$, we find $\Delta_{n,n,n,n} \to 3$ and $X_n \to 0$, and recover the results for 5D orbifold theories without BKT in~\cite{Chivukula:2001hz}. As said earlier, our calculation of the leading order of (\ref{scalarScattering}) relies only on the leading part of factors such as

\begin{equation}
\frac{s}{s-m_j^2} = 1 + \frac{m_j^2}{s-m_j^2} \, .
\end{equation}

\noindent
We will conclude the calculations of this chapter by checking that the \textit{infinite} sum of \textit{subleading} contributions is itself subleading and does not contribute to our result (\ref{leadingOrderGauge}). We modify the relevant sum in (\ref{Xsums}) by a factor $m_j^2/(s-m_j^2)$ and find

\begin{equation}
\label{surfaceSum}
\begin{split}
\sum_{j=0}^{\infty} &2^{-\delta_{j,0}} \, \tilde{\Delta}^2_{n,j,n}\; \frac{m_j^2}{s-m_j^2}\ =\ \, \tilde{\Delta}_{n,n,n,n}\, \frac{4m_n^2}{s-4 m_n^2}\\
&\quad +\; 32 \pi^2 R^2 m_n^2 \tilde{r}_c^3
N_n^4 \, \Big(\, 1 - \pi^2 R^2 m_n^2 \tilde{r}_c^2\, \Big)\, \bigg[\,\frac{m_n^2}{s-4 m_n^2} + \frac{m_n^4}{(s-4 m_n^2)^2}\,\bigg]\\
&\quad +\, \frac{8 \pi^4 R^4 m_n^4 \tilde{r}_c^5 N_n^4}{s+\frac{\sqrt{s}}{\pi R \tilde{r}_c} \tan \pi R \sqrt{s}} \,\frac{s^3-4m_n^2 s^2 + 4 m_n^4 s}{(s-4 m_n^2)^2}\ -\ 8 \pi^4 R^4 m_n^4\tilde{r}_c^5 N_n^4 \; .
\end{split}
\end{equation}

\noindent
The first two terms are obviously $\ord(m^2_n/s)$.  The third term has got poles at $\sqrt{s}=m_j$ and so a smooth high energy limit is not well defined. Taking the limit in a discrete manner, $\sqrt{s} =  (m + 1/4)/R$ with $m  \to \infty$, the third term approaches the  negative of the fourth term. Consequently, all $\ord(1)$  terms cancel and (\ref{surfaceSum}) does contribute only subleading terms.


\chapter{High-Energy Unitarity Bounds}
\label{HighEnergyUnitarityBounds}

Higher-dimensional Yang-Mills theories have got couplings of negative mass dimension, in our case $[g_5]=-1/2$, and are therefore non-renormalizable. They can only be considered as effective theories and only be trusted up to a certain energy scale. There is a limit beyond which physics of a UV complete theory must become relevant. In this chapter, we determine an upper bound on this limit by demanding perturbative unitarity of the 4D theory. Beyond this bound, one of the following three scenarios will be correct: (i) the 4D theory becomes strongly coupled, (ii) new physics sets in, or (iii) both~\cite{Chivukula:1995hr}. An upper limit on the energy translates naturally into a limit on the number of KK modes that can be produced in scattering processes.

There are different ways how unitarity of the scattering matrix can manifest itself in a theory. On one hand, there is the discussion of Martin and Froissart~\cite{Martin:1962rt, Froissart:1961ux}, who derive bounds on the \textit{total cross-section}. Total cross-sections can rise asymptotically not faster than a logarithm of the energy; a fact that we find confirmed in our calculations (\ref{leadingOrderGauge}). On the other hand, it is also possible to derive strong limits on the \textit{partial wave amplitudes} of a scattering process.

\noindent
From a naive dimensional analysis, we find the s-wave amplitude $a_0$ of a two body scattering process to increase linearly with rising energy,

\begin{equation}
a_0 \sim N_c g_5^2 \; \sqrt{s} \; .
\end{equation} 
 
\noindent
As we will discuss in greater detail below, at first approximation perturbative unitarity demands $a_0 \leq 1$. One therefore finds an upper bound $N_0$ on the number of KK modes~\cite{Chivukula:2001hz},

\begin{equation}
\label{naive}
\frac{N_0}{R} \; \lesssim \; \frac{1}{N_c \, g_5^2} \; ,
\end{equation}

\noindent
that depends on the higher-dimensional gauge coupling $g_5$, the number of colours $N_c$ and the compactification radius $R$. A more careful discussion will ultimately allow us to make statements about the $r_c$-dependence of the bound $N_0$ as well. Let us start with the standard optical theorem for transitions $|i\rangle \to |f\rangle$,

\begin{equation}
2 \operatorname{Im} T_{f i} \, = \, \sum_j T_{fj} T_{ji}^* \, ,
\end{equation} 

\noindent
where the sum is understood to include phase-space integrations for each of the individual particles in $|j\rangle$. We expand the transition amplitudes in terms of Legendre polynomials $P_l(c)$,

\begin{equation}
\label{PartialWaveExpansion}
\begin{split}
T(s,c) \, &= \, 16 \pi \sum_{l=0}^{\infty} (2l+1) P_l(c) a_l(s)\\
a_0 \, &= \, \frac{1}{32 \pi} \int_{-1}^1 dc \, T \; ,
\end{split}
\end{equation}

\noindent
where $a_l(s)$ are the partial waves and $c=\cos \theta$ the cosine of the scattering angle. Substituting (\ref{PartialWaveExpansion}) into the optical theorem, we find relations for the partial waves~\cite{Barton}. For the s-wave in particular, it reads

\begin{equation}
\label{sWaveRelation}
\operatorname{Im} [a_0]_{f i} = \sum_j \; \sigma_j \; [a_0]_{f j} \;
[a_0]^*_{j i} \; \; .
\end{equation}

\noindent
The factor $\sigma_j$ takes care of the phase-space of state $|j\rangle$. For two-particle states it has got the simple form

\begin{equation}
\begin{split}
\sigma_j \, &= \, \lambda(s,m_1^2,m_2^2)/s\\
\lambda(x,y,z) \, &= \, \sqrt{x^2+y^2+z^2 - 2(xy+yz+zx)} \; \; .
\end{split}
\end{equation}

\noindent
Absorbing the phase-space factor into the s-waves, we can rewrite (\ref{sWaveRelation}) as a matrix equation.

\begin{gather}
\label{absorb}
[\tilde{a}_0]_{ij} = \sqrt{\sigma_i \sigma_j} [ a_0 ]_{ij}\\
\label{matrixEquation}
\operatorname{Im} \tilde{a}_0 = \tilde{a}_0 \tilde{a}_0^*
\end{gather}

\noindent
Since $\tilde{a}_0$ is symmetric, it satisfies $\tilde{a}_0 \tilde{a}_0^* = (\operatorname{Re} \tilde{a}_0)^2 + (\operatorname{Im} \tilde{a}_0 )^2$. We realize that both sides of (\ref{matrixEquation}) can be diagonalized simultaneously. In general, that would have been impossible in equation (\ref{sWaveRelation}). Let $\{ \alpha_i \}$ be the set of eigenvalues of $\tilde{a}_0$ and $\alpha_{\operatorname{max}} = \operatorname{max}\{ \alpha_i \}$ the largest of them. Combining $\operatorname{Im} \alpha_i \leq |\alpha_i|$ and (\ref{matrixEquation}), we find bounds on the eigenvalues, from which the strongest one reads

\begin{equation}
\label{CCArelation}
|\alpha_{\operatorname{max}}| \; \leq \; 1 \; \; .
\end{equation}

\newpage
\noindent
The method described above is known as Coupled Channel Analysis (CCA)\footnote{Historically, relation (\ref{CCArelation}) first appears in~\cite{Lee:1977eg}. There it was used in the high energy limit, i.e. $\sigma_j=1$ and $\tilde{a}_0=a_0$, in order to derive an upper Higgs mass bound within the Standard Model. In our case, it is exactly a limit on the energy that we would like to determine, and the influence of the phase-space can no longer be neglected.}. In evaluating $\alpha_{\operatorname{max}}$, we do not merely study one isolated scattering process, but the entirety of all channels that are available at a certain energy $\sqrt{s}$. In our case, $a_0$ and therefore $\alpha_{\operatorname{max}}$ are functions of the energy. Inequality (\ref{CCArelation}) can provide us therefore with an upper KK mode bound of the effective theory, which is the aim of this chapter.

For a perfect CCA of our model, we would have to consider all scattering processes $A^a_{(n) \, \mu}  \cdots A^b_{(m) \, \nu} \; \to \; A^c_{(k) \, \rho}  \cdots A^d_{(l) \, \sigma}$ with all possible combinations of colour, polarization and KK modes. By restricting ourselves to a subset of these processes, we get a weaker but still correct bound from (\ref{CCArelation}). In what follows, we will limit our analysis to inelastic $2 \to 2$ scattering processes of longitudinal gauge bosons, $A^a_{(n) \, L}  A^a_{(n) \, L} \; \to \; A^b_{(m) \, L}  A^b_{(m) \, L}$ with $n \neq m \leq N_0$ and centre of mass energy $s=4m_{N_0}^2$. Applying the ET, i.e. replacing the longitudinal gauge bosons by the corresponding scalar modes, will simplify our calculations further. For $n,m \approx N_0$, the energy is comparable to the combined masses of the scattering particles. The use of the ET is nevertheless justified, since these channels are heavily phase-space suppressed.

There is an infinite number of tree-level diagrams contributing to the scalar scattering process $A^a_{(n) \, 5}  A^b_{(n) \, 5} \; \to \; A^c_{(m) \, 5}  A^d_{(m) \, 5}$.

\begin{align}
\SSSSblob{}{}{}{}{(n) \, p_1}{(n) \, p_2}{(m) \, k_1}{(m) \, k_2} \hspace{-2mm} &= \sum_{j=0}^{\infty} \sChannelS{}{}{}{}{(n) \, p_1}{(n) \, p_2}{(m) \, k_1}{(m) \, k_2}{\, \, \, \, (j)} \hspace{-2mm} \; + \; \cdots \nonumber \\
&\phantom{}\nonumber\\
\begin{split}
&= \sum_{j=0}^{\infty}[i T^s_{(j)}+i T^t_{(j)}+i T^u_{(j)}]\\
&= i T^s+i T^t+i T^u
\end{split}
\end{align}

\begin{equation}
\begin{split}
t &= - \frac{s}{2} +m_n^2+m_m^2+ \frac{c}{2} S_{n,m}\\
u &= - \frac{s}{2} +m_n^2+m_m^2- \frac{c}{2} S_{n,m}\\
S_{n,m} &= \sqrt{(s-4m_n^2)(s-4m_m^2)}
\end{split}
\end{equation}

\noindent
The kinematics generalizes the relations (\ref{kinematics}) with $\beta^2 = S_{n,n} / s$ and $c=\cos \theta$. The calculation of the Feynman diagrams is again straightforward.

\begin{equation}
\label{completeAmplitudes}
\begin{split}
i T^s_{(j)} &= 2^{-\delta_{j,0}} \tilde{\Delta}_{n,j,n} \tilde{\Delta}_{m,j,m} \, i g^2 f^{abe} f^{cde} \frac{u-t}{2(s-m_j^2)}\\
i T^t_{(j)} &= 2^{-\delta_{j,0}} \tilde{\Delta}^2_{n,j,m} \, i g^2 f^{ace} f^{dbe} \frac{s-u}{2(t-m_j^2)}\\
i T^u_{(j)} &= 2^{-\delta_{j,0}} \tilde{\Delta}^2_{n,j,m} \, i g^2 f^{ade} f^{bce} \frac{t-s}{2(u-m_j^2)}
\end{split}
\end{equation}

\noindent
As an aside, let us calculate the leading order of the amplitude. Using the sum (\ref{Ysums}), we find it to generalize our earlier result (\ref{leadingOrderGauge}).

\begin{equation}
\begin{split}
i&T^s+i T^t+i T^u =\\
&(\Delta_{n,n,m,m} + Y_{n,m}) i g^2 \Big[ f^{ace} f^{dbe} \frac{c^2+3}{2(c-1)} + f^{ade} f^{bce} \frac{c^2+3}{2(c+1)} \Big] + \ord \big( \frac{m_n^2}{s} \big)
\end{split}
\end{equation}

\noindent
With the help of the sum (\ref{Zsums}), a further generalization to scattering processes of arbitrary KK modes, $A^a_{(k) \, 5}  A^b_{(l) \, 5} \; \to \; A^c_{(n) \, 5}  A^d_{(m) \, 5}$, is obvious. Returning to the complete expressions (\ref{completeAmplitudes}), we now sum/average over final/initial colours and project out the s-waves\footnote{Note that zero mode exchange, $j=0$, contributes only to elastic scattering processes, see (\ref{Dnm0}) and (\ref{Dnn0}). In that case, $T^t_{(0)} \sim 1/(1-c)$ and $T^u_{(0)} \sim 1/(1+c)$, and the s-wave exhibits the same IR singularities that are familiar from standard QED. In our analysis, we will ignore these diagonal elements and set $[a_0]_{nn} \equiv 0$.}. As the energy increases, the individual contributions diverge logarithmically.

\begin{align}
[a_0]_{nm \; (j)} &= \frac{1}{32 \pi} \int_{-1}^1 dc \, [T^s_{(j)}+T^t_{(j)}+T^u_{(j)}] \nonumber \\
\begin{split}
&= - \frac{g^2 N_c}{32 \pi} \tilde{\Delta}_{n,j,m}^2 \; 2^{-\delta_{j,0}} \Big[ 1+ \frac{2(s-m_n^2- m_m^2) +m_j^2}{S_{n,m}}\\
&\qquad \times \ln \Big| \frac{s + 2(m_j^2-m_n^2 - m_m^2) - S_{n,m}}{s + 2(m_j^2-m_n^2 - m_m^2) + S_{n,m}} \Big| \Big]
\end{split}\\
\lim_{s \to \infty} [a_0]_{nm \; (j)} &= \frac{g^2 N_c}{32 \pi} \tilde{\Delta}_{n,j,m}^2 \; 2^{-\delta_{j,0}} \Big[ - 1 + 2 \ln \frac{s}{m_j^2} \Big]
\end{align}

\noindent
Summing over the complete KK tower, we find the elements of the partial wave matrix $a_0$. In the limit of vanishing $r_c$, the terms $j=m+n$ and $j=|n-m|$ dominate the sum (\ref{sWaveSum}) and we recover the result in~\cite{Chivukula:2001hz}.

\begin{align}
\label{sWaveSum}
[a_0]_{nm} \; &= \; \sum_{j=0}^{\infty} [a_0]_{nm \; (j)}\\
\lim_{\substack{s \to \infty\\ r_c \to 0}} [a_0]_{nm} &= \frac{g^2 N_c}{32 \pi} \Big[ -1 + 2 \ln \frac{s}{|n^2/R^2-m^2/R^2|} \Big]
\end{align}

\noindent
For non-vanishing BKT, summing the contributions $[a_0]_{nm \; (j)}$ analytically is challenging and we decide to perform the sum numerically. In a final step, we absorb the phase-space factor and determine the largest eigenvalue $\alpha_{\operatorname{max}}$ of the modified s-wave matrix $\tilde{a}_0$ numerically.

\begin{equation}
[\tilde{a}_0]_{nm} \; = \; \frac{S_{n,m}}{s} \, [a_0]_{nm}
\end{equation} 

\noindent
The $(N_0+1) \times (N_0+1)$ matrix $\tilde{a}_0$ is evaluated at centre of mass energy $s=4m^2_{N_0}$. $R^{-1}$ defines our fundamental scale ($R=1$ in the figures below). The maximum eigenvalue of the matrix depends therefore on three parameters, $\alpha_{\operatorname{max}}(N_0, \tilde{r}_c, g^2 N_c)$. Varying two of them, $\tilde{r}_c$ and $g^2 N_c$, condition (\ref{CCArelation}) provides us with an upper bound $N_0$ on the KK modes. The contour plot in Fig.~4.1 summarizes our numerical results.

\begin{figure}[!ht]
\begin{center}
\parbox{135mm}{
\begin{center}
\includegraphics[width=0.45\textwidth]{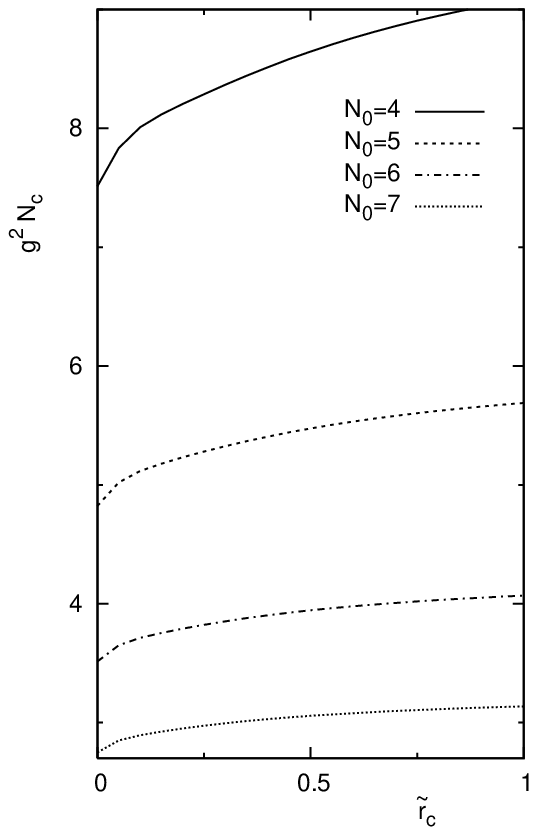}
\includegraphics[width=0.45\textwidth]{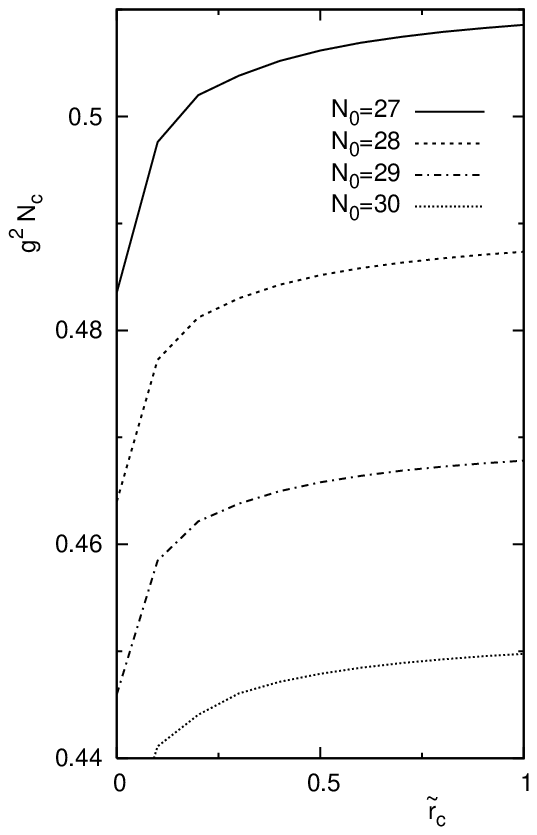}
\caption{Contours of $N_0$ in the $(\tilde{r}_c, g^2 N_c)$ plane, e.g. in the region between lines $N_0=28$ and $N_0=29$ the largest available KK mode is 28.}
\end{center}}
\end{center}
\label{KKmodeBounds}
\end{figure}

We find the bound $N_0$ to relax as the coupling $g^2 N_c$ decreases, which confirms our earlier estimate (\ref{naive}). For a theory without BKTs, $\tilde{r}_c=0$, our bound is by about a factor of two weaker than the one derived in~\cite{Chivukula:2001hz, Chivukula:2003kq}. That is a direct consequence of the phase-space corrections, that were not included in the earlier analysis. The unitarity bound relaxes as the BKT coupling $\tilde{r}_c$ increases. The overall $\tilde{r}_c$-dependence of the bound is weak and decreases further as the coupling $g^2 N_c$ falls off. This weak dependence is expected, since high energy unitarity probes distances much smaller than the compactification radius $R$. At short distances the size of a coupling at some fixed point in space becomes insignificant.
\vspace{2mm}

\noindent
We have seen that in orbifold theories without BKT the KK number is conserved. Let us consider the decay of a massive KK mode in these theories. The total mass of the decay products will exactly equal the mass of the decaying particle, e.g. $n/R = k/R + l/R$. Consequently, there is no phase-space left for the process and massive KK modes turn out to be stable.

The situation changes drastically once we allow for BKTs~\cite{Cheng:2002iz, Cheng:2002ab}. No processes are \textit{a priori} forbidden, and due to the distorted spectrum the phase-space never vanishes identically. Let $\Gamma_n$ be the decay width of a KK gauge boson $A^a_{(n) \, \mu}$ of mass $m_n$. We might expect $\Gamma_n$ to increase dramatically with rising BKT coupling $r_c$. In that case, the sensible constraint\footnote{As the CCA, this is an idea that we copy from an earlier Higgs mass bound derivation~\cite{Nachtmann}.}

\begin{equation}
\label{decayConstraint}
\frac{\Gamma_n}{2} \; \leq \; m_n
\end{equation}

\noindent
might provide us with an alternative upper KK mode bound for our effective theory. A lowest-order calculation is straightforward. Only two-body decays $A^a_{(n) \, \mu} \to A^b_{(k) \, \nu} A^c_{(l) \, \rho}$ with $k+l \leq n$ contribute. The total decay width $\Gamma_n$ is simply the sum of all partial widths.

\begin{equation}
\label{totalWidth}
\Gamma_n \; = \; \sum_{\substack{k,l=0\\ k+l \leq n}} \, \frac{\lambda (m^2_n, m^2_k, m^2_l)}{16 \pi m_n^3} |T_{(n,k,l)}|^2
\end{equation}

\noindent
We sum/average the squared amplitudes $T_{(n,k,l)}$ over final/initial colours and polarizations, and finally find

\begin{equation}
\begin{split}
|T_{(n,k,l)}|^2 = &\frac{g^2 N_c}{3} \pi^2 R^2 \tilde{r}_c^3 N_n N_k N_l \Delta_{n,k,l}\\
&\times \, \Big[ m_n^4 + m_k^4 + m_l^4 + 10\big( m_n^2 m_k^2 + m_k^2 m_l^2 +m_l^2 m_n^2 \big) \Big] \; .
\end{split}
\end{equation}

\noindent
Note that $\Delta_{n,k,l}$ does not appear quadratically, since one of the coefficients cancels against contributions from the polarization vector sum. Interestingly, decays into zero modes do not contribute, although the channels are clearly kinematically allowed. Due to (\ref{Dnm0}), amplitudes $T_{(n,k,0)}$ with $0 \leq k < n$ vanish identically. Finally, we perform the finite sum (\ref{totalWidth}) numerically.

\begin{figure}[!hb]
\begin{center}
\parbox{95mm}{
\begin{center}
\includegraphics[width=0.65\textwidth]{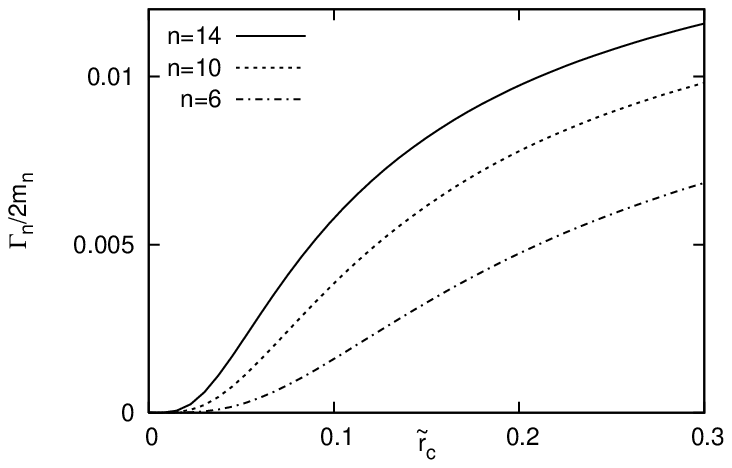}\\ \vspace{3mm}
\includegraphics[width=0.65\textwidth]{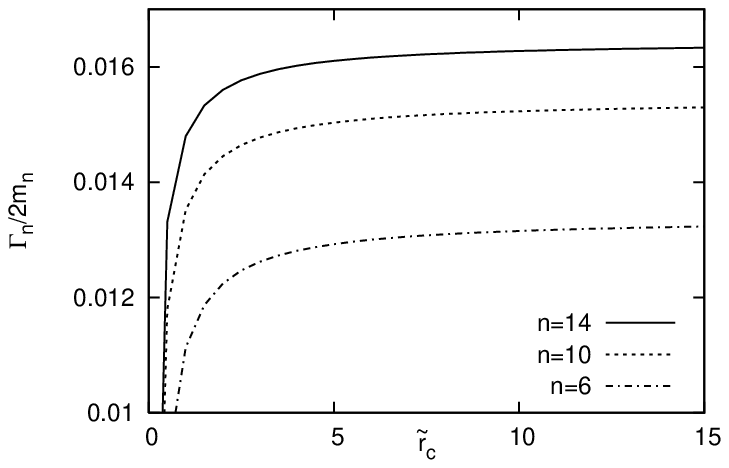}
\caption{Decay width mass ratio $\Gamma_n/2 m_n$ with respect to BKT coefficient $\tilde{r}_c$.}
\end{center}}
\end{center}
\label{widthrc}
\end{figure}

\begin{figure}[!hb]
\begin{center}
\parbox{95mm}{
\begin{center}
\includegraphics[width=0.65\textwidth]{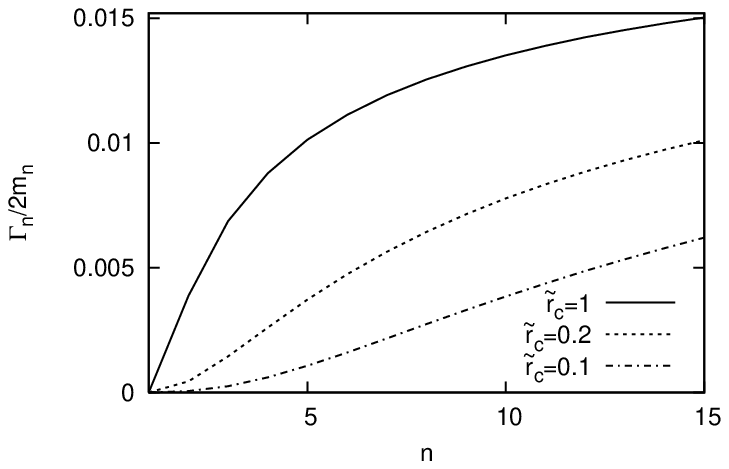}\\ \vspace{3mm}
\includegraphics[width=0.65\textwidth]{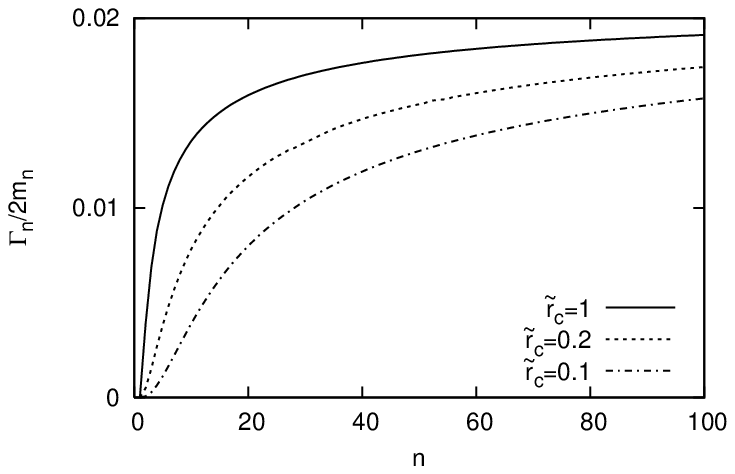}
\caption{Decay width mass ratio $\Gamma_n/2 m_n$ with respect to KK mode $n$.}
\end{center}}
\end{center}
\label{widthn}
\end{figure}

Fig.~4.2 shows a plot of the decay width mass ratio with respect to the BKT coefficient $\tilde{r}_c$. It confirms the importance of BKTs for the decays, that we mentioned earlier on. In the limit $\tilde{r}_c \to 0$, all KK modes become stable. In Fig.~4.3 we plot the same ratio with respect to the KK mode number $n$, i.e. effectively the mass $m_n \approx n/R$. Heavy modes decay comparatively faster than light ones. What is important in both plots is that they saturate far from the critical value $\Gamma_n/(2 m_n) = 1$. Consequently, constraint (\ref{decayConstraint}) can provide us with no additional bound, as we had initially hoped.


\chapter{Conclusions}
\label{Conclusions}

We have studied 5D Yang-Mills theories compactified on an $S^1/\Zb_2$ orbifold with kinetic terms localized at the fixed points. These terms, the so-called BKT, are strictly necessary in any consistent quantum field theoretic discussion of orbifold theories.
\vspace{3mm}

\noindent
We have presented a novel quantization method for these theories and derived powerful Ward and Slavnov-Taylor identities. In the process, we have developed the concept of functional differentiation on orbifolds. Note that we did not regularize the BKT in any form; no thick branes~\cite{delAguila:2003bh}, discretized extra dimension~\cite{delAguila:2004xd} or $\epsilon$-regularized wave functions~\cite{Muck:2004br} have been used.

We have presented a general all-order proof of the Generalized Equivalence Theorem. Despite the difficult spectrum, we have succeeded in deriving highly non-trivial sum rules, that lie at the heart of the high energy unitarity cancellations. We developed a novel Coupled Channel Analysis which can be used at finite centre-of-mass energies. Our calculations have shown that limits derived from high energy unitarity are not sensitive to the size of localized terms. On the other hand, we have demonstrated the great importance of these terms for the decay of the KK modes.
\vspace{3mm}

\noindent
If extra dimensions exist there are likely to be more than just one, as indicated by ADD scenarios and string theory. Many of the applications presented in the Introduction make use of two extra space dimensions. The author is currently working on a generalization of the concepts presented in this thesis to 6D orbifold theories. The single BKT considered in this study is only the first of a whole series of higher-order correction terms. A complete brane renormalization discussion~\cite{Goldberger:2001tn, pLewandowski:2001q, delAguila:2003bh, Oliver:2003cy} will be more involved. Finally, it will be necessary to introduce fermions~\cite{Papavassiliou:2001be} in our model.


\begin{appendix}
\chapter{Feynman rules}
\label{FeynmanRules}

The classical part of the Lagrangian (\ref{Lag5DYM}) gives rise to the following two-point functions for the KK mass eigenstates.

\begin{align*}
\BB{a \; \mu}{b \; \nu}{(n) \; p}
&\parbox{110mm}{
\begin{equation}
\Gamma^{a b}_{\mu \nu} (p_{(n)})\ =\ i \delta^{ab} \big[ 
-g_{\mu \nu}\, \big( p^2 - m_n^2\big)\: +\: 
p_{\mu} p_{\nu}\, \big]
\end{equation}}\\
\BS{a \; \mu}{b}{(n) \; p}
&\parbox{110mm}{
\begin{equation}
\label{mixing}
\Gamma^{a b}_{\mu 5} (p_{(n)})\ =\ \delta^{ab} m_n p_{\mu}
\end{equation}}\\
\SSLL{a}{b}{(n) \; p}
&\parbox{110mm}{
\begin{equation}
\Gamma^{a b}_{5  5} (p_{(n)})\ =\ i \delta^{ab} p^2
\end{equation}}\\
\end{align*}

\noindent
Note that (\ref{mixing}) represents a non-vanishing mixing between vector and scalar modes. After quantization of the theory by including terms $\Lag_{5D \GF}$ and $\Lag_{5D \FP}$ in the Lagrangian, these mixing terms cancel and the two-point functions take on the form given below.

\begin{align*}
\BB{a \; \mu}{b \; \nu}{(n) \; p}
&\parbox{110mm}{
\begin{equation}
\Gamma^{ab}_{\mu \nu} (p_{(n)})\ =\ i \delta^{ab} \big[ 
-g_{\mu \nu}\, ( p^2 - m_n^2)\: +\: 
\big( 1 - \mbox{$\frac{1}{\xi}$}\big)\, p_{\mu} p_{\nu}\,\big]
\end{equation}}\\
\SSLL{a}{b}{(n) \; p}
&\parbox{110mm}{
\begin{equation}
\Gamma^{ab}_{55} (p_{(n)})\ =\ i \delta^{ab} \big( p^2\: -\: \xi
m_n^2 \big)
\end{equation}}\\
\GG{a}{b}{(n) \; p}
&\parbox{110mm}{
\begin{equation}
\Gamma^{ab}_{\bar{c} c} (p_{(n)})\ =\ -i \delta^{ab} \big( p^2\: -\:
\xi m_n^2\, \big)
\end{equation}}
\end{align*}

\noindent
The corresponding propagators for vector, scalar and ghost KK modes read as follows.

\begin{align}
D^{ab}_{\mu \nu} (p_{(n)})\ &=\ \frac{i
\delta^{ab}}{p^2-m_n^2+i\epsilon}\ \bigg[ -\,g_{\mu \nu}\: +\: (1- \xi)\,
\frac{p_{\mu} p_{\nu}}{p^2-\xi m_n^2}\, \bigg]\\ 
D^{ab}_{55} (p_{(n)})\ &=\
\frac{i \delta^{ab}}{p^2- \xi m_n^2 +i \epsilon}\\
D^{ab}_{\overline{c} c} (p_{(n)})\ &=\ \frac{-i \delta^{ab}}{p^2- \xi
m_n^2 +i \epsilon}
\end{align}

\noindent
Based on the results of Appendix~\ref{Products}, we can derive the fundamental interactions of the KK modes. In the expressions below, we work with the dimensionless coupling $g=g_5 / \sqrt{2 \pi R}$. The explicit forms of the coefficients $\Delta_{k,l,n}$, $\tilde{\Delta}_{k,l,n}$, $\Delta_{k,l,n,m}$ and $\tilde{\Delta}_{k,l,n,m}$ are given in equations (\ref{DeltaThree}) and (\ref{DeltaFour}). 

\begin{align*}
\BBBeast{a \; \mu}{b \; \nu}{c \; \rho}{(n) \; k}{(m) \; p}{(l) \; q}
&\parbox{110mm}{
\begin{equation}
\begin{split}
\Gamma^{abc}_{\mu \nu \rho} &(k_{(n)},p_{(m)},q_{(l)})\ =\ \\ &g f^{abc}
\sqrt{2}^{\; -1-\delta_{n,0}-\delta_{m,0}-\delta_{l,0}} \,
\Delta_{n,m,l}\\ &\times \big[g_{\mu \nu}(k-p)_{\rho}+g_{\rho
\mu}(q-k)_{\nu}+g_{\nu \rho}(p-q)_{\mu} \big]
\end{split}
\end{equation}}\\ &\phantom{}\\
\BBSeast{b \; \mu}{c \; \nu}{a}{(m) \; p}{(l) \; q}{(n) \; k}
&\parbox{110mm}{
\begin{equation}
\begin{split}
\Gamma^{abc}_{5 \mu \nu} &(k_{(n)},p_{(m)},q_{(l)})\ =\ i g f^{abc} \,
g_{\mu \nu}\\ & \times \big[ m_l \sqrt{2}^{\; -1-\delta_{m,0}}
\tilde{\Delta}_{n,m,l} - m_m \sqrt{2}^{\; -1-\delta_{l,0}}
\tilde{\Delta}_{n,l,m} \big]
\end{split}
\end{equation}}\\ &\phantom{}\\
\SSBeast{b}{c}{a \; \mu}{(m) \; p}{(l) \; q}{(n) \; k}
&\parbox{110mm}{
\begin{equation}
\begin{split}
\Gamma^{abc}_{\mu 5 5} (&k_{(n)},p_{(m)},q_{(l)})\ =\\
&g f^{abc} \sqrt{2}^{\; -1-\delta_{n,0}} \tilde{\Delta}_{l,n,m} (q-p)_{\mu}
\end{split}
\end{equation}}\\
\end{align*}

\begin{align*}
\BBBB{a \; \mu}{b \; \nu}{c \; \rho}{d \; \sigma}{(n) \; k}{(m) \;
p}{(l) \; q}{(k) \; r} &\parbox{110mm}{
\begin{equation}
\begin{split}
\Gamma^{abcd}_{\mu \nu \rho \sigma} (&k_{(n)},p_{(m)},q_{(l)},r_{(k)})
\ =\ \\ i g^2 &\Delta_{n, m, l, k} \sqrt{2}^{\;
 -2-\delta_{n,0}-\delta_{m,0}-\delta_{l,0}-\delta_{k,0}}\\ \times
 \big[ &f^{abe} f^{cde} (g_{\mu \sigma} g_{\nu \rho} - g_{\mu \rho}
 g_{\nu \sigma}) +\\ &f^{ace} f^{bde} (g_{\mu \sigma} g_{\nu \rho} -
 g_{\mu \nu} g_{\rho \sigma})+\\ &f^{ade} f^{bce} (g_{\mu \rho} g_{\nu
 \sigma} - g_{\mu \nu} g_{\rho \sigma}) \big]
\end{split}\\
\end{equation}}\\
\BBSS{a \; \mu}{b \; \nu}{c}{d}{(n) \; k}{(m) \; p}{(l) \; q}{(k) \; r}
&\parbox{110mm}{
\begin{equation}
\begin{split}
\Gamma^{abcd}_{\mu \nu 5 5} &(k_{(n)},p_{(m)},q_{(l)},r_{(k)})\ =\ i g^2
\sqrt{2}^{\; -2 -\delta_{n,0}-\delta_{m,0}}\\ &\times
\tilde{\Delta}_{n,m,l,k} \, g_{\mu \nu} \big[ f^{ace} f^{bde} +
f^{ade} f^{bce} \big]
\end{split}
\end{equation}}
\end{align*}

\begin{align*}
\GGBeast{a}{b}{c \; \mu}{(n) \; k}{(m) \; p}{(l) \; q}
&\parbox{110mm}{
\begin{equation}
\begin{split}
\Gamma^{abc}_{\overline{c} c \mu} &(k_{(n)}, p_{(m)}, q_{(l)})\ =\ \\ &-g
f^{abc} \sqrt{2}^{\; -1-\delta_{n,0}-\delta_{m,0}-\delta_{l,0}} \,
\Delta_{n,m,l} \, k^{\mu}
\end{split}
\end{equation}}\\ &\phantom{}\\
\GGSeast{a}{b}{c}{(n) \; k}{(m) \; p}{(l) \; q}
&\parbox{110mm}{
\begin{equation}
\begin{split}
\Gamma^{abc}_{\overline{c} c 5} &(k_{(n)}, p_{(m)}, q_{(l)})\ =\ \\ &i g
f^{abc} \xi m_l \sqrt{2}^{\; -1-\delta_{n,0}-\delta_{m,0}} \,
\Delta_{n,m,l}
\end{split}
\end{equation}}
\end{align*}


\chapter{Mass eigenmode expansion}
\label{MassEigenmodeExpansion}

In Section~\ref{5DYM}, we discussed in detail the compactification of 5D Yang-Mills theories with BKT on an orbifold $S^1/\Zb_2$. In what follows, we will derive the analytic form of the complete set of orthonormal functions $f_n$ and $g_n$ that were used in expansion (\ref{componentExpansion}) of the components of the higher-dimensional gauge field. Our discussion in the first part of this appendix will be based on an orbifold theory with a single BKT at $y=0$, as used in the calculations of Chapters~\ref{GET} and~\ref{HighEnergyUnitarityBounds}. In a second part we will then focus on a theory with two BKTs, in order to demonstrate the full generality of our approach.


\section{Brane kinetic term at $y=0$}
\label{1BKT}

We have already seen that in our quantization scheme, the full Lagrangian~(\ref{Lag5DYM}) is proportional to a common factor $[1+r_c \delta(y)]$. One can think of this factor as part of the integration measure of the full quantized action and not the Lagrangian itself, cf. definitions of the generating functionals (\ref{defZ}) and (\ref{defGamma}). It therefore makes sense to demand our set of functions to be orthonormal with respect to this measure.

\begin{equation}
\label{orthonormality}
\begin{split}
\int_{- \pi R}^{\pi R} dy \; \big[ 1 + r_c \delta(y) \big] \, f_n(y) f_m(y) &= \delta_{n,m}\\
\int_{- \pi R}^{\pi R} dy \; \big[ 1 + r_c \delta(y) \big] \, g_n(y) g_m(y) &= \delta_{n,m}
\end{split}
\end{equation}

\noindent
The $\Zb_2$ orbifold symmetry allows for two different parities, and we introduce the $f_n$ as even and the $g_n$ as odd functions in $y$. Both of them are $2 \pi R$-periodic.

\begin{equation}
\label{parity}
\begin{split}
f_n(y) &= f_n(-y)\\
g_n(y) &= - g_n(-y)
\end{split}
\end{equation}

\noindent
Let us now have a closer look at the kinetic part of the 5D Lagrangian~(\ref{Lag5DYM}). The last terms in line two and three of~(\ref{5DkineticPart}) are the undesirable mixing terms that we would like to see cancelled. We come to them in a minute.

\begin{equation}
\label{5DkineticPart}
\begin{split}
\Lag_{\rm 5D \YM} (x,y) \supset &\big[ 1 + r_c \delta (y) \big] \Big[ - \frac{1}{4} (\partial_\mu A^a_\nu - \partial_\nu A^a_\mu)(\partial_\mu A^{a \nu} - \partial^\nu A^{a \mu})\\
&- \frac{1}{2} (\partial_5 A^a_\mu)(\partial^5 A^{a \mu}) - \frac{1}{2} (\partial_\mu A^a_5)(\partial^\mu A^{a \, 5}) + (\partial_5 A^a_\mu)(\partial^\mu A^{a \, 5})\\
&-\frac{1}{2 \xi} (\partial^\mu A^a_\mu)^2 - \frac{\xi}{2} (\partial_5 A^a_5)^2 + (\partial^\mu A^a_\mu)(\partial_5 A^a_5) \Big]
\end{split}
\end{equation}

\noindent
After compactification, the 4D effective Lagrangian should describe a theory in which each individual KK mode is quantized in the conventional $R_\xi$ gauge.

\begin{equation}
\label{4DkineticPart}
\begin{split}
\Lag_{\operatorname{eff}} (x) \supset &- \frac{1}{4} (\partial_\mu A^a_{(n) \nu} - \partial_\nu A^a_{(n) \mu})(\partial_\mu A^{a \nu}_{(n)} - \partial^\nu A^{a \mu}_{(n)})\\
&+ \frac{1}{2} m_n^2 A^a_{(n) \mu} A^{a \, \mu}_{(n)} - \frac{1}{2 \xi} (\partial_\mu A^{a \, \mu}_{(n)})^2\\
&+ \frac{1}{2}(\partial_\mu A^a_{(n) 5})(\partial^\mu A^a_{(n) 5}) - \frac{\xi}{2} m_n^2 (A^a_{(n) 5})^2
\end{split}
\end{equation}

\noindent
Given the orthonormality introduced earlier, the functions $f_n$ and $g_n$ must therefore satisfy the following simple wave equations.

\begin{equation}
\label{waveEquations}
\begin{split}
\big[ \partial_5^2 + m_n^2 \big] f_n(y) &= 0\\ 
\big[ \partial_5^2 + m_n^2 \big] g_n(y) &= 0
\end{split}
\end{equation}

\noindent
Instead of comparing the Lagrangians (\ref{5DkineticPart}) and (\ref{4DkineticPart}) directly, we can base our arguments on the equations of motion derived from them, as done in~\cite{Carena:2002me}. The final wave equations are the same. Note that they hold in the entire interval $(-\pi R, \pi R]$. The $\delta(y)$-term in the equivalent equation in~\cite{Carena:2002me} stems from the particular gauge choice, $A^a_5(x,y)=0$, made there. This choice is of course not suitable for a discussion of the GET.

Let us now return to the mixing terms that appear in~(\ref{5DkineticPart}). Integration by parts of the first term is not problematic, since the fields are understood to vanish at 4D infinity, and we find

\begin{equation}
\Lag_{\rm 5D \YM} (x,y) \supset \big[ 1 + r_c \delta (y) \big] \big[ (\partial_5 \partial^\mu A^a_\mu ) A^a_5 + (\partial^\mu A^a_\mu )( \partial_5 A^a_5 ) \big] \; .
\end{equation}

\noindent
The fields of the first term will be expanded in terms of $\partial_5 f_n$ and $g_n$, whereas the expansion of the second term will be based on $f_n$ and $\partial_5 g_n$ instead. The functions $\partial_5 f_n$ and $\partial_5 g_n$ are odd/even, and have consequently an expansion in terms of the $g_n$ and $f_n$ respectively. We assume this expansion to be particularly simple, $\partial_5 f_n = c_1 g_n$ and $\partial_5 g_n = c_2 f_n$. The 4D effective theory will be absent of any mixing terms, if and only if $c_1=-c_2$. Combining this statement with (\ref{waveEquations}), we arrive at

\begin{equation}
\label{lastConstraint}
\begin{split}
\partial_5 \, f_n(y) \; &= \; - m_n g_n(y)\\
\partial_5 \, g_n(y) \; &= \; m_n f_n(y) \; .
\end{split}
\end{equation}

\noindent
The combined constraints (\ref{orthonormality}), (\ref{parity}), (\ref{waveEquations}) and (\ref{lastConstraint}) are sufficient to uniquely specify the analytic form of the functions $f_n$ and $g_n$: Assume $f_n(y)$ to be a solution of wave equation (\ref{waveEquations}) in the interval $0 < y \leq \pi R$. It must therefore be a linear combination of sine and cosine. The constraints (\ref{parity}) and (\ref{lastConstraint}) then fix the form of $f_n(y)$ for negative arguments and the form of $g_n(y)$ in the entire definition interval. This information is reflected in the ansatz below.

\begin{align}
f_n(y) =& \begin{cases}
- A_n \sin m_n y + B_n \cos m_n y \; &\textrm{for} \quad -\pi R < y \leq 0\\
\; \; \; A_n \sin m_n y + B_n \cos m_n y \; &\textrm{for}\qquad 0 < y \leq \pi R
\end{cases}\\[3mm]
g_n(y) =& \begin{cases}
B_n \sin m_n y + A_n \cos m_n y \; &\textrm{for} \quad -\pi R < y < 0\\
B_n \sin m_n y - A_n \cos m_n y \; &\textrm{for}\qquad 0 < y < \pi R\\
0 &\textrm{for} \qquad y=0 \; \; \textrm{or} \; \; y=\pi R
\end{cases}
\end{align}

\noindent
Outside the interval $(-\pi R, \pi R]$, both functions are $2 \pi R$-periodic by construction. We \textit{define} the derivative of $f_n(y)$ and $g_n(y)$ at a point $y$ as the average of left and right derivatives at this point. For both $f_n(y)$ and $g_n(y)$, left and right derivatives are identical at every single point, which includes $g_n(y)$ at $y=0$. The only exception is $f_n(y)$ at $y=0$. Left and right derivatives differ in sign, and we find $\partial_5 f_n(0)=0$ in accordance with $g_n(0)=0$. The upshot of this paragraph is that both functions and their derivatives are unambiguously defined on $\Rb$.

From the normalization of both functions, i.e. $n=m \neq 0$ in (\ref{orthonormality}), and the orthogonality of the $f_n(y)$, i.e. $n \neq m =0$ in (\ref{orthonormality}), we determine the three free parameters $A_n$, $B_n$ and $m_n$ of our ansatz.

\begin{gather}
\int_{- \pi R}^{\pi R} dy \; \big[ 1 + r_c \delta(y) \big] f_n^2 = 1\nonumber \\
\int_{- \pi R}^{\pi R} dy \; \big[ 1 + r_c \delta(y) \big] f_n = 0\\
\int_{- \pi R}^{\pi R} dy \; g_n^2 = 1\nonumber
\end{gather}

\begin{figure}[!ht]
\begin{center}
\parbox{120mm}{
\begin{center}
\includegraphics[width=0.7\textwidth]{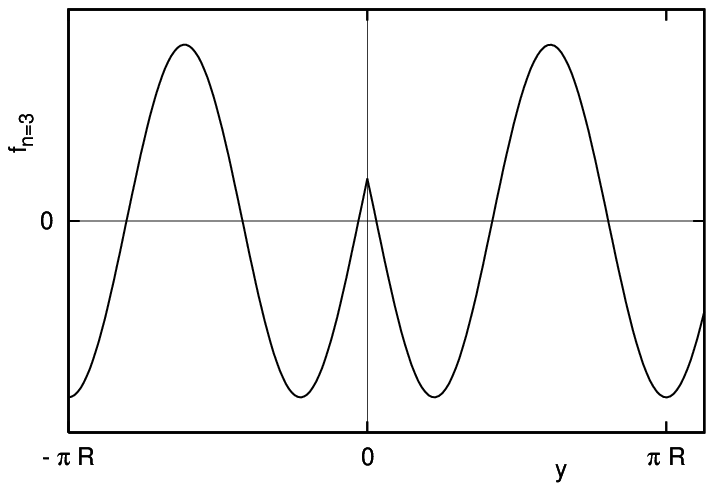}\\
\includegraphics[width=0.7\textwidth]{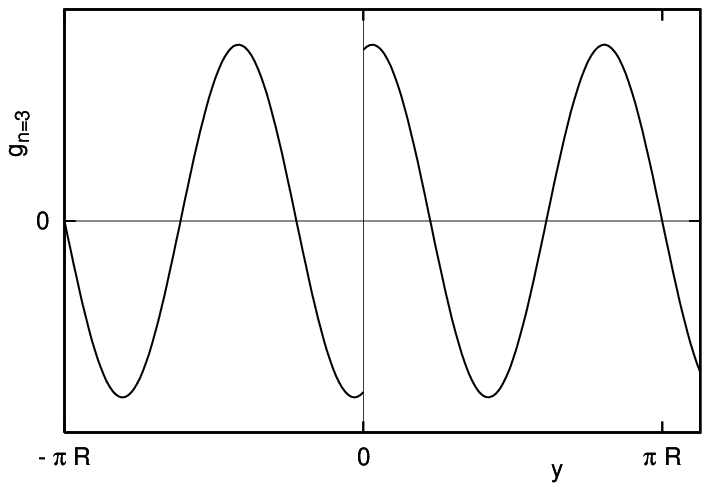}
\caption{Mass eigenstate wave functions $f_n(y)$ and $g_n(y)$ with $n=3$ for an orbifold theory with a single BKT at $y=0$.}
\end{center}}
\end{center}
\end{figure}

\noindent
The final expression for the complete set of orthonormal functions reads

\begin{align}
\label{fnpure}
f_n(y) =& \frac{N_n}{\sqrt{2^{\delta_{n,0}} \pi R}} \times \begin{cases}
\cos m_n y + \frac{1}{2} m_n r_c \sin m_n y \; &\textrm{for} \quad -\pi R < y \leq 0\\
\cos m_n y - \frac{1}{2} m_n r_c \sin m_n y \; &\textrm{for}\qquad 0 < y \leq \pi R
\end{cases}\\[3mm]
\label{gnpure}
g_n(y) =& \frac{N_n}{\sqrt{\pi R}} \times \begin{cases}
\sin m_n y - \frac{1}{2} m_n r_c \cos m_n y \; &\textrm{for} \quad -\pi R < y < 0\\
\sin m_n y + \frac{1}{2} m_n r_c \cos m_n y \; &\textrm{for}\qquad 0 < y \leq \pi R\\
0 &\textrm{for} \qquad y=0
\end{cases}
\end{align}

\noindent
where

\begin{equation}
\label{Nn}
N_n^{-2} = 1 + \tilde{r}_c + \pi^2 R^2 \tilde{r}_c^2 m_n^2 \qquad \textrm{with} \quad \tilde{r}_c = \frac{r_c}{2 \pi R} > 0
\end{equation}

\noindent
is a normalization constant. The spectrum $m_n$ is defined by the following transcendental equation.

\begin{equation}
\label{spectrumAppendix}
\frac{m_n r_c}{2} = - \tan m_n \pi R
\end{equation}

\noindent
Using it in (\ref{fnpure}) and (\ref{gnpure}), we can rewrite the $f_n$ and $g_n$ as shifted sines and cosines and arrive at (\ref{fn}) and (\ref{gn}). As expected, we recover the standard Fourier expansion in the limit of vanishing BKTs.

\begin{equation}
\begin{split}
\lim_{r_c \to 0} \, f_n (y) \; &= \; \frac{1}{\sqrt{2^{\delta_{n,0}} \pi R}} \cos \frac{ny}{R}\\ 
\lim_{r_c \to 0} \, g_n (y) \; &= \; \frac{1}{\sqrt{\pi R}} \sin \frac{ny}{R}\\ 
\end{split}
\end{equation}

\noindent
When discussing functional differentiation on $S^1/\Zb_2$ in Chapter~\ref{WardSlavnovTaylor}, we came across the delta function $\delta(y; r_c)$. Its defining relation reads 

\begin{equation}
\label{testfunction}
\int_{- \pi R}^{\pi R} dy \; [1+r_c \delta(y)] \, h(y) \delta(y-y'; \, r_c) = h(y') \; ,
\end{equation}

\noindent
where we recognise the familiar factor in the integration measure. At $y'=0$, definition (\ref{testfunction}) is non-trivial only for even test functions $h(y)$. Substituting an even expansion for $\delta(y; r_c)$ as well as $h(y)$ into (\ref{testfunction}), we can derive an explicit form of the delta function.

\begin{align}
\label{deltaExpansion}
\delta(y; r_c) \; &= \; \sum_{n=0}^\infty \; \frac{N_n^2}{2^{\delta_{n,0}} \pi R} \, \frac{\cos m_n (y \pm \pi R)}{\cos m_n \pi R}\\
&= \;\begin{cases}
1/r_c \; &\textrm{for} \quad y = 0\\
0 \; &\textrm{for} \quad - \pi R < y < 0 \;\, \textrm{or} \;\, 0 < y \leq \pi R
\end{cases} \nonumber
\end{align}

\noindent
In the limit $r_c \to 0$, we find the Fourier expansion of the delta function on the circle. It is the discrete equivalent of $\delta(x) = \int dp/2 \pi \; \exp(ixp)$ on $\Rb$.

\begin{equation}
\lim_{r_c \to 0} \delta (y; \, r_c) = \delta (y) = \sum_{n=0}^{\infty} \, \frac{1}{2^{\delta_{n,0}}\,\pi R} \cos \frac{ny}{R}
\end{equation}

\noindent
So far, the \textit{completeness} of our set of orthonormal functions $f_n$ and $g_n$ has not entered our discussion. With the help of (\ref{deltaExpansion}), we are now in the position to check it explicitly.

\begin{equation}
\label{completeness}
\delta (y_1-y_2; \, r_c) = \sum_{n=0}^{\infty}\,\big[f_n (y_1) \, f_n (y_2) + g_n (y_1) \, g_n(y_2)\big]
\end{equation}


\section{Brane kinetic terms at $y=0$ and $y=\pi R$}
\label{2BKT}

We include this section in order to demonstrate the flexibility of our approach. It can easily accommodate for a second BKT\footnote{Even a generalization to 6D orbifold theories with kinetic terms at orbifold fixed points or \textit{lines} seems to be straightforward.} at the fixed point $y=\pi R$. All that is needed is a replacement of the factors which include the delta functions.

\begin{equation}
\label{replacement}
\big[1 + r_c \delta(y)\big] \; \to \; \big[1 + r_c \delta(y) + r_c \delta(y- \pi R) \big]
\end{equation}

\noindent
Our orbifold theory with two BKTs is defined by Lagrangian (\ref{Lag5DYM}) and quantization terms (\ref{5DGF}) and (\ref{5DFP}), subject to replacements (\ref{replacement}). In what follows, we will sketch the derivation of the complete set of functions used for compactification. The new orthonormality conditions read

\begin{equation}
\begin{split}
\int_{- \pi R}^{\pi R} dy \; \big[ 1 + r_c \delta(y) + r_c \delta(y- \pi R) \big] f_k(y) f_l(y) &= \delta_{k,l}\\
\int_{- \pi R}^{\pi R} dy \; \big[ 1 + r_c \delta(y) + r_c \delta(y- \pi R) \big] g_k(y) g_l(y) &= \delta_{k,l} \; .
\end{split}
\end{equation}

\noindent
The remaining constraints (\ref{parity}), (\ref{waveEquations}), and (\ref{lastConstraint}) as well as the entire ansatz for $f_n$ and $g_n$ remain unchanged. The three parameters $A_n$, $B_n$ and $m_n$ can now be determined from  

\begin{gather}
\int_{- \pi R}^{\pi R} dy \; \big[ 1 + r_c \delta(y) + r_c \delta(y- \pi R) \big] f_n^2 = 1\nonumber \\
\int_{- \pi R}^{\pi R} dy \; \big[ 1 + r_c \delta(y) + r_c \delta(y- \pi R) \big] f_n = 0\\
\int_{- \pi R}^{\pi R} dy \; g_n^2 = 1 \; . \nonumber
\end{gather}

\begin{figure}[!ht]
\begin{center}
\parbox{120mm}{
\begin{center}
\includegraphics[width=0.7\textwidth]{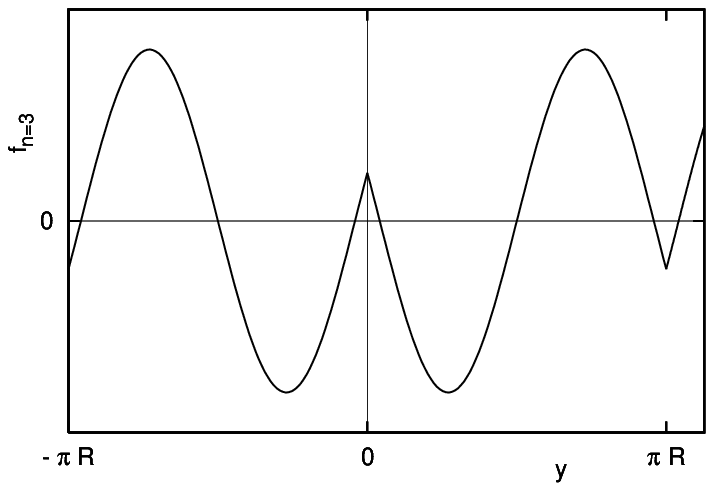}\\
\includegraphics[width=0.7\textwidth]{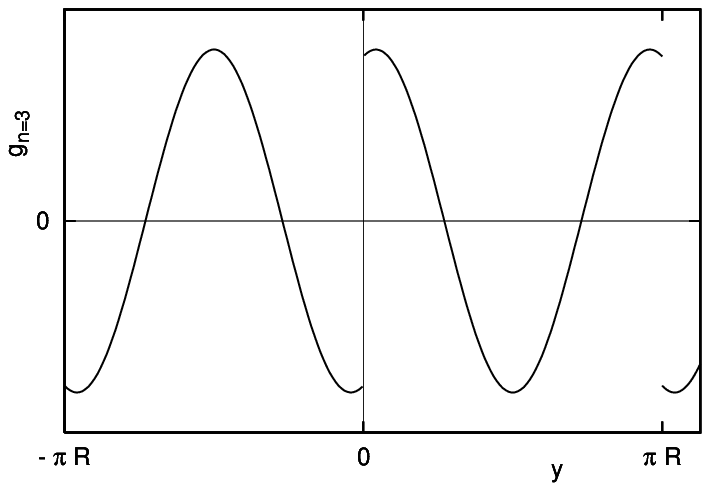}
\caption{Mass eigenstate wave functions $f_n(y)$ and $g_n(y)$ with $n=3$ for an orbifold theory with two BKTs at $y=0$ and $y=\pi R$.}
\end{center}}
\end{center}
\end{figure}

\noindent
The analytic expressions for $f_n(y)$ and $g_n(y)$ are still of the form (\ref{fnpure}) and (\ref{gnpure}), where the normalization constant is now given by

\begin{equation}
\label{2BKTnormalization}
\begin{split}
N_n^{-2} = 1 + 4 \tilde{r}_c &+ \frac{1+2 \tilde{r}_c}{(1-\tilde{r}_c \pi R m_n)^2} + \frac{1+2 \tilde{r}_c}{(1+\tilde{r}_c \pi R m_n)^2}\\
&- \frac{1+3\tilde{r}_c}{1-\tilde{r}_c \pi R m_n} - \frac{1+3\tilde{r}_c}{1+\tilde{r}_c \pi R m_n} \; .
\end{split}
\end{equation}

\noindent
The spectrum $m_n$ is again defined by a transcendental equation and can be expressed in the following factorised form.

\begin{equation}
\label{2BKTspectrum}
m_n\, \big[ \tan \frac{m_n \pi R}{2} + \frac{m_n r_c}{2}\, \big]\, \big[\, \cot \frac{m_n \pi R}{2} - \frac{m_n r_c}{2}\, \big] = 0
\end{equation}

\noindent
The solutions of the equation can be read off from Fig. 2.1. The masses $m_n$ are slightly lighter than the corresponding ones in the one-BKT spectrum. They vary in the limit $(n-1)/R < m_n \leq n/R$ for $\infty > r_c \geq 0$. Interestingly and unlike in theories with a single BKT, the mass gap between zero mode and lightest KK mode does vanish as $r_c \to \infty$. We can again express the set of orthonormal functions in terms of shifted sines and cosines. Due to the new spectrum, we now have to distinguish between even and odd KK numbers.

\begin{align}
f_n(y) =& \frac{N_n}{\sqrt{2^{\delta_{n,0}} \pi R} \; \cos (m_n \pi R / 2)} \times \begin{cases}
\; \; \; \cos m_n ( y \pm \pi R / 2) \; &\textrm{for even} \; \; n\\
\pm \sin m_n ( y \pm \pi R / 2) \; &\textrm{for odd} \; \; n
\end{cases}\\[3mm]
g_n(y) =& \frac{N_n}{\sqrt{\pi R} \; \sin (m_n \pi R / 2)} \times \begin{cases}
\; \; \; \sin m_n ( y \pm \pi R / 2) \; &\textrm{for even} \; \; n\\
\mp \cos m_n ( y \pm \pi R / 2) \; &\textrm{for odd} \; \; n
\end{cases}
\end{align}

\noindent
The different signs in the above expressions refer to the negative and positive definition intervals respectively. The distinction between even and odd KK modes as well as the more complex analytic expressions for normalization constants (\ref{2BKTnormalization}) and mass spectrum (\ref{2BKTspectrum}) would make calculations equivalent to the ones in Appendices~\ref{Products} and~\ref{summation} more tiresome. But we hope to have convincingly shown, that the underlying concepts of the quantization scheme, the GET proof as well as the discussion of the high energy unitarity bounds remain unaffected.


\chapter{Products on $S^1/\Zb_2$}
\label{Products}

When compactifying higher-dimensional theories, we repeatedly come across products of functions that need to be integrated over the compact dimension $y$. Take for example $A^c_\mu(x,y) \theta^b(x,y)$ in (\ref{gauge5D}) which leads to (\ref{gaugeKK4D}). The product of these two even functions is again even and can be expanded in terms of the functions $f_n(y)$. Compactification boils down to finding the coefficients of this expansion in terms of $A^c_{(n) \, \mu}(x)$ and $\theta^b_{(n)}(x)$.

There are five products, i.e. five combinations of even and odd functions, that reappear in our calculations. In this appendix, we determine the coefficients of their expansion. The appendix is thought of as a detailed repository, that should be helpful when following our calculations in the main chapters.
 
We will use the orthonormal basis (\ref{fn}) and (\ref{gn}) to expand even and odd functions respectively.

\begin{alignat}{2}
F_+(y)&=F_+(-y) & G_-(y)&=-G_-(-y)\nonumber \\
F_+(y)&=\sum_{n=0}^{\infty} F_{+(n)} f_n(y) &\hspace{2cm} G_-(y)&=\sum_{n=1}^{\infty} G_{-(n)} g_n(y)\nonumber 
\end{alignat}

\noindent
Let us start by considering the product of two even functions, as in the example mentioned above. Using the orthonormality (\ref{orthonormality}), the coefficients of the expansion can readily be expressed as below. 

\begin{align}
F_+(y)G_+(y)&=\sum_{n=0}^{\infty} [F_+*G_+]_{(n)} f_n(y) \nonumber \\
\label{ee}
\begin{split}
[F_+*G_+]_{(n)}&=\int_{-\pi R}^{\pi R} dy \, [1+r_c \delta(y)] F_+(y) G_+(y) f_n(y)\\
&=\sum_{k,l=0}^{\infty} F_{+(k)} G_{+(l)} \int_{-\pi R}^{\pi R} dy \, [1+r_c \delta(y)] f_k(y) f_l(y) f_n(y)
\end{split}
\end{align}

\noindent
We are left with an integral over a product of three basis functions. In a first step, we rewrite this product as a simple sum of cosines.

\begin{align}
\begin{split}
\int_{-\pi R}^{\pi R} dy \, &[1+r_c \delta(y)] f_k f_l f_n =\\
&\frac{N_k N_l N_n}{\sqrt{2^{\delta_{k,0}+\delta_{l,0}+\delta_{n,0}} \pi^3 R^3}} \Big[ r_c + \frac{2}{\cos m_k \pi R \, \cos m_l \pi R \, \cos m_n \pi R}\\
&\times \int_0^{\pi R}dy \, \cos m_k(y-\pi R) \cos m_l(y-\pi R) \cos m_n(y-\pi R) \Big]
\end{split} \nonumber \\
& \hspace{3.2cm} =  \frac{\Delta_{k, l, n}}{2 \sqrt{2^{\delta_{k,0}+\delta_{l,0}+\delta_{n,0}} \pi R}}
\end{align}

\begin{equation}
\begin{split}
\Delta_{k,l,n} = &\Delta(m_k, m_l, m_n) + \Delta(-m_k, m_l, m_n)\\
&+ \Delta(m_k, -m_l, m_n) + \Delta(m_k, m_l, -m_n)
\end{split}
\end{equation}

\noindent
The integration is now trivial. The final result distinguishes between the two cases of vanishing and non-vanishing period.

\begin{align}
\label{DeltaDrei}
\Delta&(m_k,m_l,m_n) = N_k N_l N_n \Bigg[ \frac{\int_0^{\pi R} dy \, \cos (m_k+m_l+m_n)(y-\pi R)}{\pi R \, \cos m_k \pi R \, \cos m_l \pi R \, \cos m_n \pi R} + \tilde{r}_c \Bigg] \\[3mm]
&= 
\begin{cases}
N_k N_l N_n \pi^2 R^2 \tilde{r}_c^3 \, {\displaystyle \frac{m_k m_l m_n}{m_k+m_l+m_n}}\\
\textrm{for} \quad m_k+m_l+m_n \neq 0\\[4mm]
N_k N_l N_n\Big([(1+\pi^2 R^2 \tilde{r}_c^2 m_k^2)(1+\pi^2 R^2 \tilde{r}_c^2 m_l^2)(1+\pi^2 R^2 \tilde{r}_c^2 m_n^2)]^{\frac{1}{2}}+\tilde{r}_c \Big)\\
\textrm{for} \quad m_k+m_l+m_n = 0
\end{cases} \nonumber
\end{align}

\noindent
We therefore arrive at an explicit expression for the coefficients (\ref{ee}).

\begin{equation}
[F_+*G_+]_{(n)}=\sum_{k,l=0}^{\infty} F_{+(k)} G_{+(l)} \, \frac{\Delta_{k,l,n}}{2 \sqrt{2^{\delta_{k,0}+\delta_{l,0}+\delta_{n,0}} \pi R}}
\end{equation}

\noindent
Along the same lines, we derive the coefficients of four further products.

\begin{equation}
\begin{split}
F_+(y) G_-(y) &= \sum_{n=1}^{\infty} [F_+*G_-]_{(n)}g_n(y)\\
[F_+*G_-]_{(n)}&=\sum_{k,l=0}^{\infty} F_{+(k)} G_{-(l)} \, \frac{\tilde{\Delta}_{l,k,n}}{2 \sqrt{2^{\delta_{k,0}} \pi R}}
\end{split}
\end{equation}

\begin{equation}
\begin{split}
F_-(y) G_-(y) &= \sum_{n=0}^{\infty} [F_-*G_-]_{(n)}f_n(y)\\
[F_-*G_-]_{(n)}&=\sum_{k,l=0}^{\infty} F_{-(k)} G_{-(l)} \, \frac{\tilde{\Delta}_{k,n,l}}{2 \sqrt{2^{\delta_{n,0}} \pi R}}
\end{split}
\end{equation}

\begin{equation}
\begin{split}
F_+(y) G_+(y) H_+(y) &= \sum_{n=0}^{\infty} [F_+*G_+*H_+]_{(n)} f_{(n)}(y)\\
[F_+*G_+*H_+]_{(n)}&=\sum_{k,l,m=0}^{\infty} F_{+(k)} G_{+(l)} H_{+(m)} \, \frac{\Delta_{k,l,m,n}}{4 \pi R \sqrt{2^{\delta_{k,0}+\delta_{l,0}+\delta_{m,0}+\delta_{n,0}}}}
\end{split}
\end{equation}

\begin{equation}
\begin{split}
F_-(y) G_-(y) H_+(y) &= \sum_{n=0}^{\infty} [F_-*G_-*H_+]_{(n)} f_{(n)}(y)\\
[F_-*G_-*H_+]_{(n)}&=\sum_{k,l,m=0}^{\infty} F_{-(k)} G_{-(l)} H_{+(m)} \, \frac{\tilde{\Delta}_{n,m,k,l}}{4 \pi R \sqrt{2^{\delta_{m,0}+\delta_{n,0}}}}
\end{split}
\end{equation}

\noindent
The factors appearing in the coefficients are linear combinations of expressions (\ref{DeltaDrei}) and (\ref{DeltaVier}). The arguments of $\Delta(m_k, m_l, m_n)$ and $\Delta(m_k, m_l, m_n,m_m)$ can either be elements of the spectrum or of its negative.

\begin{equation}
\label{DeltaThree}
\begin{split}
\Delta_{k,l,n} = &\; \Delta(m_k, m_l, m_n) + \Delta(-m_k, m_l, m_n)\\
&+ \Delta(m_k, -m_l, m_n) + \Delta(m_k, m_l, -m_n)\\[1mm]
\tilde{\Delta}_{k,n,l} = &- \Delta(m_k, m_l, m_n) - \Delta(m_k, m_l, -m_n)\\
&+ \Delta(-m_k, m_l, m_n) + \Delta(m_k, -m_l, m_n)
\end{split}
\end{equation}

\begin{equation}
\label{DeltaFour}
\begin{split}
\Delta_{k,l,n,m} = &\; \Delta(m_k,m_l,m_n,m_m) + \Delta(-m_k,m_l,m_n,m_m)\\
&+ \Delta(m_k,-m_l,m_n,m_m) + \Delta(m_k,m_l,-m_n,m_n)\\
&+ \Delta(m_k,m_l,m_n,-m_m) + \Delta(-m_k,-m_l,m_n,m_m)\\ 
&+ \Delta(-m_k,m_l,-m_n,m_m) + \Delta(-m_k,m_l,m_n,-m_m)\\[2mm]
\tilde{\Delta}_{k,l,n,m} = &- \Delta(m_k,m_l,m_n,m_m) - \Delta(-m_k,m_l,m_n,m_m)\\
&- \Delta(m_k,-m_l,m_n,m_m) - \Delta(-m_k,-m_l,m_n,m_m)\\
&+ \Delta(m_k,m_l,-m_n,m_m) + \Delta(m_k,m_l,m_n,-m_m)\\ 
&+ \Delta(-m_k,m_l,-m_n,m_m) + \Delta(-m_k,m_l,m_n,-m_m)
\end{split}
\end{equation}

\noindent
$\Delta_{k,l,n}$ and $\Delta_{k,l,n,m}$ are symmetric in all their indices. $\tilde{\Delta}_{k,l,n}$ is symmetric under permutation of the first and third index, whereas $\tilde{\Delta}_{k,l,n,m}$ is symmetric under the exchange of its first two as well as its last two indices. If one of the indices is zero, the following relations hold.

\begin{equation}
\begin{split}
\Delta_{k,l,n,0} &= \frac{2}{\sqrt{1+\tilde{r}_c}} \Delta_{k,l,n}\\
\tilde{\Delta}_{0,k,l,n} &= \frac{2}{\sqrt{1+\tilde{r}_c}} \tilde{\Delta}_{l,k,n}\\
\tilde{\Delta}_{k,l,n,0} &= 0
\end{split}
\end{equation}

\begin{align}
\Delta&(m_k,m_l,m_n,m_m) \nonumber \\
\label{DeltaVier}
&= N_k N_l N_n N_m \, \Bigg[\frac{\int_0^{\pi R} dy \, \cos (m_k+m_l+m_n+m_m)(y-\pi R)}{\pi R \, \cos m_k \pi R \, \cos m_l \pi R \, \cos m_n \pi R \, \cos m_m \pi R} + \tilde{r}_c \Bigg]\\[3mm]
&=
\begin{cases}
N_k N_l N_n N_m \; \pi^2 R^2 \tilde{r}_c^3 \, {\displaystyle \frac{m_l m_n m_m + m_k m_n m_m + m_k m_l m_m + m_k m_l m_n}{m_k+m_l+m_n+m_m}}\\
\textrm{for} \quad m_k+m_l+m_n+m_m \neq 0\\[3mm]
N_k N_l N_n N_m \; [(1+\pi^2 R^2 \tilde{r}_c^2 m_k^2)(1+\pi^2 R^2 \tilde{r}_c^2 m_l^2)\\
\hspace{2cm} \times (1+\pi^2 R^2 \tilde{r}_c^2 m_n^2)(1+\pi^2 R^2 \tilde{r}_c^2 m_m^2)]^{\frac{1}{2}}+\tilde{r}_c\\
\textrm{for} \quad m_k+m_l+m_n+m_m = 0
\end{cases} \nonumber
\end{align}

\noindent
In the case that no combination of the masses vanishes, i.e. $m_k + m_l \neq m_n$ for all permutations of the indices, expressions (\ref{DeltaThree}) simplify as follows.

\begin{equation}
\begin{split}
\Delta_{k, l, n} &= N_k N_l N_n \, \pi^2 R^2 \tilde{r}_c^3 \, \frac{8 m_k^2 m_l^2 m_n^2}{m_k^4+m_l^4+m_n^4-2(m_k^2 m_l^2 + m_l^2 m_n^2 + m_n^2 m_k^2)}\\
\tilde{\Delta}_{k, n, l} &= N_k N_l N_n \, \pi^2 R^2 \tilde{r}_c^3 \, \frac{4 m_k m_l m_n^2(m_k^2+m_l^2-m_n^2)}{m_k^4+m_l^4+m_n^4-2(m_k^2 m_l^2 + m_l^2 m_n^2 + m_n^2 m_k^2)}
\end{split}
\end{equation}

\begin{equation}
\label{Dnm0}
\begin{split}
\Delta_{n,n,j} &= N_n^2 N_j \pi^2 R^2 \tilde{r}_c^3 \, \frac{8 m_n^4}{m_j^2-4 m_n^2}\\
\tilde{\Delta}_{n,n,j} &= N_n^2 N_j \pi^2 R^2 \tilde{r}_c^3 \, \frac{4 m_j m_n^3}{m_j^2-4m_n^2}\\
\tilde{\Delta}_{n,j,n} &= - N_n^2 N_j \pi^2 R^2 \tilde{r}_c^3 \, 4 m_n^2 \frac{m_j^2-2m_n^2}{m_j^2-4 m_n^2}\\
\Delta_{n,m,0} &=\tilde{\Delta}_{n,0,m} =0 \qquad \text{for} \quad n \neq m
\end{split}
\end{equation}

\noindent
On the other hand, if there are vanishing mass combinations, the lower cases in (\ref{DeltaDrei}) and (\ref{DeltaVier}) become relevant and we find 

\begin{equation}
\label{Dnn0}
\begin{split}
\Delta_{n,n,0} = \tilde{\Delta}_{n,0,n} &= \frac{2 N_n^2}{\sqrt{1+\tilde{r}_c}} [1+\tilde{r}_c+\pi^2 R^2 \tilde{r}_c^2 m_n^2]\\
\tilde{\Delta}_{n,n,0} &= 0
\end{split}
\end{equation}

\begin{equation}
\begin{split}
\Delta_{n, n, n, n} &= 3 \tilde{\Delta}_{n,n,n,n} = 3 N_n^4 [\tilde{r}_c(1-\pi^2 R^2 \tilde{r}_c^2 m_n^2)+(1+\pi^2 R^2 \tilde{r}_c^2 m_n^2)^2]\\
\Delta_{n,n,m,m} &= 2 N_n^2 N_m^2 [ 1 + \tilde{r}_c + (1- \tilde{r}_c) \tilde{r}_c^2 \pi^2 R^2 (m_n^2+m_m^2)+ \tilde{r}_c^4 \pi^4 R^4 m_n^2 m_m^2 ]\\
\Delta_{0,0,0,0}&= \frac{2}{\sqrt{1+\tilde{r}_c}} \Delta_{0,0,0}= \frac{8}{1+\tilde{r}_c}\\
\tilde{\Delta}_{0,0,0,0}&=\tilde{\Delta}_{0,0,0}=0 \; .
\end{split}
\end{equation}

\noindent
Let us finally comment on the limit $r_c \to 0$. Consider for example $\Delta(m_k, m_l, m_n)$. The limit is trivial for cases in which the (in)equalities between the masses do not change as $r_c \to 0$, e.g. $m_n + m_0 = m_n$ and $n/R + 0 = n/R$. The first line in (\ref{DeltaDrei}) vanishes, whereas the second approaches one. Now consider the case that a relation between the masses becomes fulfilled as $r_c \to 0$, for example $m_k + m_l \neq m_{k+l}$ but $k/R + l/R = (k+l)/R$. Taking the limit of the factor $r_c^3/(m_k+m_l-m_{k+l})$ in (\ref{DeltaDrei}) carefully,
\begin{gather}
\lim_{r_c \to 0} \frac{d^3m_n}{d r_c^3}=-\frac{n^3}{4 \pi R^4} + \ord(n) \nonumber \\
\lim_{r_c \to 0} \frac{r_c^3}{m_k+m_l-m_{k+l}}=\lim_{r_c \to 0} \frac{6}{\frac{d^3 m_k}{d r_c^3}+\frac{d^3 m_l}{d r_c^3}-\frac{d^3 m_{k+l}}{d r_c^3}}=\frac{8 \pi R^4}{kl(k+l)} \; , \nonumber
\end{gather}

\noindent
we find
\begin{equation}
\lim_{r_c \to 0} \Delta(m_k,m_l,-m_{k+l})=1 \; .
\end{equation}

\noindent
We can see that our complex delta expressions (\ref{DeltaDrei}) and (\ref{DeltaVier}) reduce to simple Kronecker symbols.
\begin{equation}
\label{DeltaLimit}
\begin{split}
\lim_{r_c \to 0} \; \Delta(m_k,m_l,m_n) &= \delta_{k+l+n,0}\\
\lim_{r_c \to 0} \; \Delta(m_k,m_l,m_n,m_m) &= \delta_{k+l+n+m,0}
\end{split}
\end{equation}

\noindent
In particular, we find $\Delta_{k,l,n} \to \delta_{k,l,n}$, $\tilde{\Delta}_{k,l,n} \to \tilde{\delta}_{k,l,n}$ etc., and we reproduce the results for orbifold theories without BKT~\cite{Muck:2001yv}. Coefficients such as $\delta_{k,l,n}$ are defined as simple combinations of Kronecker deltas, see (\ref{DeltaThree}) and (\ref{DeltaFour}), and imply selection rules for the couplings of orbifold theories without BKT.


\chapter{Summation over KK modes}
\label{summation}

In our calculations of scattering amplitudes in Chapter~\ref{GET}, we encountered repeatedly infinite sums over intermediate KK modes. Take for example the elastic scattering of two vector KK modes (\ref{gaugeScattering}), where we have to calculate the infinite sum 

\begin{equation*}
\sum_{j=0}^{\infty} \; 2^{-\delta_{j,0}}\,\Delta^2_{n,n,j} \, .
\end{equation*}

\noindent
Considering (\ref{Dnm0}), the problem reduces to finding the sum

\begin{equation}
\label{realisticSum}
\sum_{j=1}^{\infty} \; \frac{N_j^2}{(m_j^2-4m_n^2)^2} \; .
\end{equation}

\noindent
Although we do not have explicit expressions for the masses $m_j$, we are able to calculate sums of this kind. In this appendix we will show how. We will develop a complex analysis summation technique, first discussed in~\cite{Bhattacharyya:2002vf}. Let us start with the simple sum

\begin{equation}
\label{simpleSum}
\sum_{j=1}^\infty \; N_j^2 \; ,
\end{equation}

\noindent
where $N_j$ is the normalization constant derived in (\ref{Nn}). In the region around the mass spectrum, $|z-m_j| < \epsilon$, we have

\begin{equation*}
z+\frac{2}{r_c} \, \tan \pi R z \approx (z-m_j) \Big[ 1+ \frac{1}{\tilde{r}_c \cos^2 \pi R z} \Big] \; .
\end{equation*}
 
\noindent
The relation remains correct for the negative spectrum, and we introduce the convention $-m_j \equiv m_{-j}$. Note that the LHS has the form of the spectrum (\ref{spectrumAppendix}). In the next step, we integrate the inverse of the above expression in the complex plane, where the contour $C_n$ encircles the entire spectrum. Although the integrands are two different functions, their residua are identical and the integrals on LHS and RHS are equal.

\begin{equation*}
\lim_{n \to \infty} \, \oint_{C_n} dz \, \Big[ z+ \frac{2}{r_c} \tan \pi R z \Big]^{-1} = \lim_{n \to \infty} \, \oint_{C_n} dz \, \Big[ 1+ \frac{1}{\tilde{r}_c \cos^2 \pi R z} \Big]^{-1} \, \sum_{j=-\infty}^{\infty} \frac{1}{z-m_j}
\end{equation*}

\noindent
The contours $C_n$ are circles $z_n = (n+1/4)/R \: \exp (i \theta)$ in the complex plane. Their radius approaches infinity in a discrete manner, $n \to \infty$, which ensures that none of the poles of the integrands lie on the contour. We evaluate the LHS by explicitly performing the $\theta$-integration. On the RHS, we apply the residue theorem.

\begin{equation*}
2 \pi i = 2 \pi i \, \sum_{j=-\infty}^{\infty} \, \Big[ 1+ \frac{1}{\tilde{r}_c} + \tilde{r}_c \pi^2 R^2 m_j^2 \Big]^{-1}
\end{equation*}

\noindent
On the RHS, we have eliminated the cosine with the help of the spectrum (\ref{spectrumAppendix}). After a slight rearrangement, we arrive at our desired sum~(\ref{simpleSum}).

\begin{equation*}
\frac{1}{2 \tilde{r}_c(1+\tilde{r}_c)} = \sum_{j=1}^{\infty} \; N_j^2
\end{equation*}

\noindent
Very good, but it is (\ref{realisticSum}) that we need for the calculation of our scattering amplitude. The terms in the sum differ by a factor $(m_j^2-4 m_n^2)^{-2}$. We modify our summation technique by multiplying the integrand under the complex integration by a factor $(z^2-4m_n^2)^{-2}$. On the LHS, the new factor suppresses the integrand as we approach the contour at infinity. The explicit $\theta$-integration gets us $0$ instead of the former $2 \pi i$. On the RHS, two new double poles at $z= \pm 2 m_n$ appear in addition to the infinite number of poles of the spectrum. Taking their residua into account we find

\begin{equation}
\begin{split}
\sum_{j=1}^{\infty} \; &N_j^2 \, \frac{1}{(m_j^2-4m_n^2)^2}=\\
&\frac{\tilde{r}_c(1-\pi^2 R^2 m_n^2 \tilde{r}_c^2)+(1+\pi^2 R^2 m_n^2 \tilde{r}_c^2)^2}{64 \pi^4 R^4 m_n^8 \tilde{r}_c^6}-\frac{1+\tilde{r}_c-\pi^2 R^2 m_n^2 \tilde{r}_c^2}{32 \pi^2 R^2 m_n^6 \tilde{r}_c^3 (1+ \tilde{r}_c)} \; .
\end{split}
\end{equation}

\noindent
With a modifying factor $z^2 \, (z^2-4m_n^2)^{-2}$ the LHS integrand is again suppressed and the LHS $\theta$-integral vanishes. Taking care of the new residua, we find the sum

\begin{equation}
\sum_{j=1}^{\infty} \; N_j^2 \, \frac{m_j^2}{(m_j^2-4m_n^2)^2} =\frac{\tilde{r}_c(1-\pi^2 R^2 m_n^2 \tilde{r}_c^2)+(1+\pi^2 R^2 m_n^2 \tilde{r}_c^2)^2}{16 \pi^4 R^4 m_n^6 \tilde{r}_c^6} \; .
\end{equation}

\noindent
A factor $z^4 \, (z^2-4m_n^2)^{-2}$ approaches one at infinity, and the $\theta$-integration remains unchanged, i.e. $2 \pi i$ on the LHS as in the calculation of the original sum (\ref{simpleSum}). Again, there are two additional residua on the right, and we get

\begin{equation}
\begin{split}
\sum_{j=1}^{\infty} \; N_j^2 \, \frac{m_j^4}{(m_j^2-4m_n^2)^2} &= \frac{(1+\pi^2 R^2 m_n^2 \tilde{r}_c^2)(1+ \tilde{r}_c + \pi^2 R^2 m_n^2 \tilde{r}_c^2)}{4 \pi^2 R^2 m_n^4 \tilde{r}_c^6}\\
&= \frac{\Delta_{n,n,n,n}+ \frac{3}{4} \, X_n}{12 N_n^4 \pi^4 R^4 m_n^4 \tilde{r}_c^6} \; .
\end{split}
\end{equation}

\noindent
Using the form (\ref{Dnm0}) of the delta expressions, combinations of the three sums above get us 

\begin{equation}
\label{Xsums}
\begin{split}
\sum_{j=0}^{\infty} 2^{-\delta_{j,0}} \Delta_{n,n,j}^2 &= \Delta_{n,n,n,n}\\
\sum_{j=1}^{\infty} \tilde{\Delta}_{n,n,j}^2 &= \tilde{\Delta}_{n,n,n,n}\\
\sum_{j=0}^{\infty} 2^{-\delta_{j,0}} \Delta_{n,n,j} \tilde{\Delta}_{n,j,n} &= \tilde{\Delta}_{n,n,n,n}\\
\sum_{j=0}^{\infty} 2^{-\delta_{j,0}} \tilde{\Delta}_{n,j,n}^2 &= \Delta_{n,n,n,n} + X_n\\[3mm]
X_n &= 8\,N_n^4\, \pi^2 R^2 \tilde{r}_c^3 \, m_n^2 \; .
\end{split}
\end{equation}

\noindent
Along the same lines, we can calculate the more general sums below, which are used in the calculation of the high energy unitarity bounds in Chapter~\ref{HighEnergyUnitarityBounds}.

\begin{equation}
\label{Ysums}
\begin{split}
\sum_{j=0}^{\infty} 2^{-\delta_{j,0}} \Delta_{n,j,m}^2 &= \Delta_{n,n,m,m}\\
\sum_{j=0}^{\infty} 2^{-\delta_{j,0}} \tilde{\Delta}_{n,j,n} \tilde{\Delta}_{m,j,m} &= \Delta_{n,n,m,m} + Y_{n,m}\\
\sum_{j=0}^{\infty} 2^{-\delta_{j,0}} \tilde{\Delta}_{n,j,m}^2 &= \Delta_{n,n,m,m} + Y_{n,m}\\[3mm]
Y_{n,m} &= 4\,N_n^2 N_m^2\, \pi^2 R^2 \tilde{r}_c^3\, (m_n^2 + m_m^2)
\end{split}
\end{equation}

\begin{equation} 
\label{Zsums}
\begin{split}
\sum_{j=0}^{\infty} 2^{-\delta_{j,0}} &\tilde{\Delta}_{k,j,l} \tilde{\Delta}_{n,j,m} = \Delta_{k,l,n,m} + Z_{k,l,n,m}\\[2mm]
Z_{k,l,n,m} &= 2 \, N_k N_l N_n N_m \, \pi^2 R^2 \tilde{r}_c^3\, (m_k^2 + m_l^2 + m_n^2 + m_m^2)
\end{split}
\end{equation}

\noindent
The simple relations (\ref{Xsums}) to (\ref{Zsums}) and their laborious derivation leave the impression that there must be a more elegant way of deriving them.

\end{appendix}


\bibliographystyle{utphys}
\bibliography{thesis.bib}

\end{document}